\documentclass[a4paper,aps,prd,11pt]{article}
\pdfoutput=1
\usepackage[utf8]{inputenc}
\usepackage{jcappub}
\usepackage{natbib}
\setcitestyle{square,numbers}
\usepackage{xcolor}
\usepackage{float}
\usepackage{multirow,array,longtable}
\usepackage{bigstrut}
\usepackage{arydshln} 
\usepackage[makeroom]{cancel}
\usepackage{amsfonts}
\usepackage{dcolumn}
\usepackage{bm}
\usepackage{graphicx}
\usepackage{amssymb,amsmath}
\usepackage{multirow}
\usepackage{slashed,xfrac}
\usepackage[makeroom]{cancel}
\usepackage{rotating}
\usepackage{color,url}
\usepackage[colorlinks=true
,urlcolor=blue
,anchorcolor=blue
,citecolor=blue
,filecolor=blue
,linkcolor=blue
,menucolor=blue
,linktocpage=true
,pdfproducer=medialab
,pdfa=true
]{hyperref}
\usepackage{epsfig,psfrag,rotating,soul}
\usepackage{rotfloat}
\usepackage[normalem]{ulem}
\usepackage{framed,fancybox,xcolor,xfrac,geometry}
\usepackage{hhline}
\renewcommand\eqref[1]{Eq.~(\ref{#1})}
\newcommand\eqrefs[2]{Eqs.~(\ref{#1})-(\ref{#2})}
\newcommand\figref[1]{Fig.~\ref{#1}}
\newcommand\figrefs[2]{Figs.~\ref{#1}-\ref{#2}}
\newcommand\tabref[1]{Table~\ref{#1}}

\newcommand\secref[1]{Section~\ref{#1}}

\evensidemargin \oddsidemargin
\allowdisplaybreaks

\def\gY{g'}
\def\gYc{g^{'\,2}}

\def\mw{m_{\rm W}}
\def\mz{m_{\rm Z}}
\def\mh{m_{\rm H}}

\def\vev{{\it v}}

\newcommand{\nn}{\nonumber}
\newcommand{\be}{\begin{equation}}
\newcommand{\ee}{\end{equation}}
\newcommand{\bear}{\begin{eqnarray}}
\newcommand{\eear}{\end{eqnarray}}
\newcommand{\mL}{\mathcal{L}}
\newcommand{\mO}{\mathcal{O}}

\newcommand{\mF}{\mathcal{F}}

\newcommand{\mV}{\mathcal{V}}
\def\1loop{one-loop}

\def\amp{\mathcal{A}}

\usepackage{soul}
\usepackage{lineno} 

\title{Double Higgs production at TeV $\bf{e^+e^-}$ colliders with Effective Field Theories: sensitivity to BSM  Higgs couplings}
\author[a]{D. Domenech,}
\author[a]{M. J.  Herrero,}
\author[b]{ R. A.  Morales}
\author[a]{and M. Ramos}
\affiliation[a]{Departamento de F\'{\i}sica Te\'orica and Instituto de F\'{\i}sica Te\'orica, IFT-UAM/CSIC,\\
Universidad Aut\'onoma de Madrid, Cantoblanco, 28049 Madrid, Spain}
\affiliation[b]{IFLP, CONICET - Dpto. de F\'{\i}sica, Universidad Nacional de La Plata, \\ 
C.C. 67, 1900 La Plata, Argentina}
\emailAdd{jose.domenech@estudiante.uam.es}
\emailAdd{maria.herrero@uam.es}
\emailAdd{roberto.morales@fisica.unlp.edu.ar}
\emailAdd{maria.pestanadaluz@uam.es}

\abstract{In this work we study the production of two Higgs bosons at the two planned electron positron colliders with energies at the TeV domain, CLIC and ILC,  within the context of Effective Field Theories (EFTs) to describe beyond the Standard Model Higgs Physics.  We focus first on the case of the Higgs Effective Field Theory (HEFT) and next we compare with the case of the Standard Model Effective Field Theory (SMEFT).  The predictions for double Higgs production in both EFTs are first presented for the most relevant subprocess participating in the total process of our interest,  $e^+e^- \to HH \nu \bar \nu$,  which is the scattering of two gauge bosons,  $WW \to HH$,  also called $WW$ fusion.  The predictions for the cross section
$ \sigma(W_X W_Y\to HH)$ as a function of the subprocess energy are analyzed in full detail for the two EFTs, for all the polarization channels with longitudinal and transverse modes $XY=LL, TT,LT,TL$, and for the most relevant effective operators in both cases.  We will demonstrate that in the HEFT case,  the total cross section can be fully understood in terms of the $LL$  contribution and this in turn is dominated at these TeV energies mainly by two HEFT coefficients.  By doing the matching between the two EFTs at the level of the scattering amplitude for the subprocess, we will be able to find the correspondence of the leading coefficients in the HEFT and  the SMEFT. In the final part of this work we will then explore the sensitivity to these two most relevant HEFT coefficients, at CLIC (3 TeV,  5 ab$^{-1}$) and ILC (1 TeV, 8 ab$^{-1}$).  We will then conclude on the accessible region of these two parameters by studying the predicted rates at these two $e^+e^-$ colliders for the final state 
$b \bar b b \bar b \nu \bar \nu$ leading to characteristic signals with 4 bottom jets and missing energy.}
 
\begin{document}
\begin{flushright}
	IFT-UAM/CSIC-22-79 
\end{flushright}
\maketitle

\section{Introduction}
\begin{figure}[h!]
\begin{center}
\includegraphics[scale=0.3]{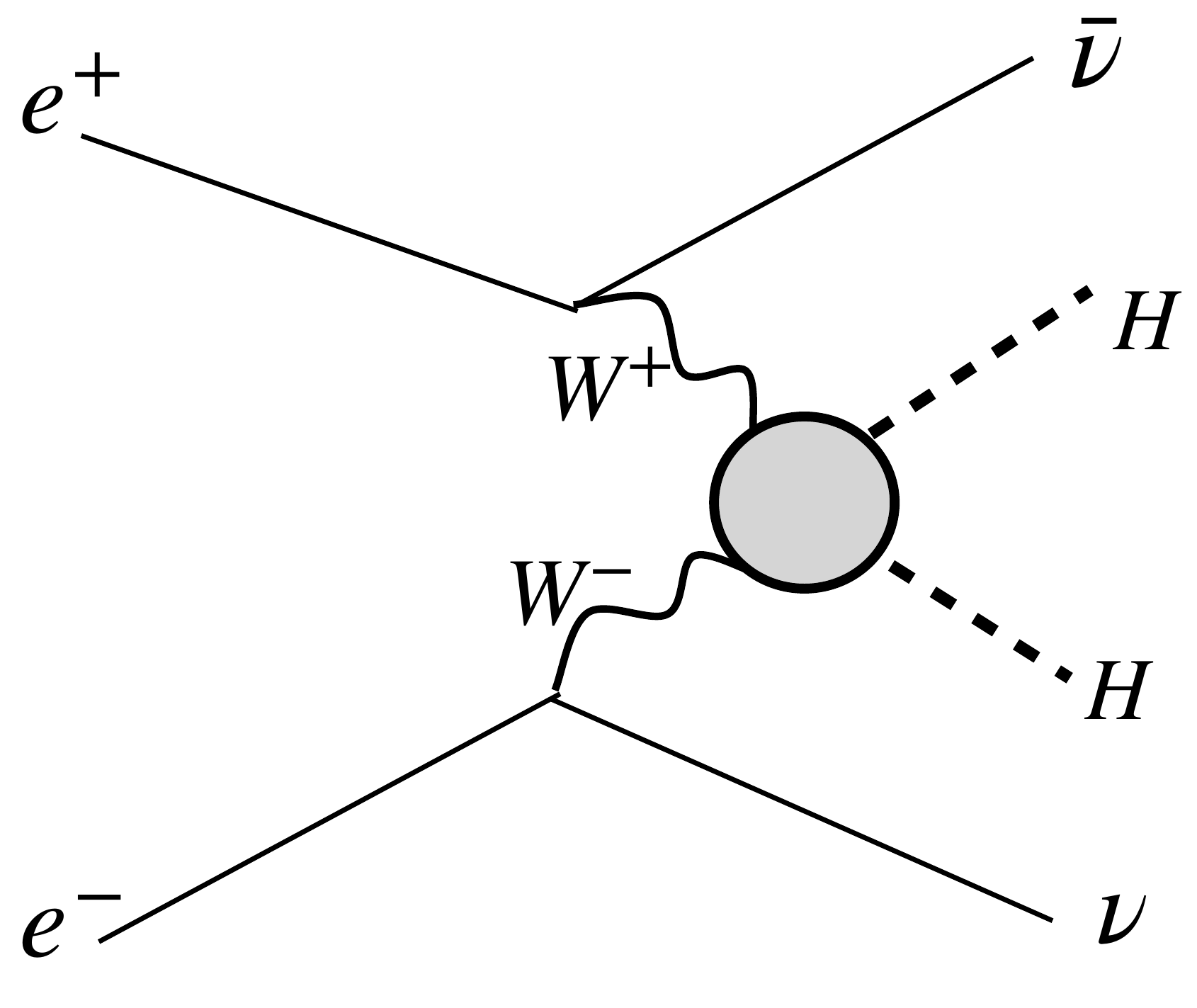} 
\caption{Double Higgs production at $e^+e^-$ via $WW$ fusion}
\label{plot-ee}
\end{center}
\end{figure}

Double Higgs production at high energy $e^+e^-$ colliders in the TeV region is one of the most promising mechanisms to test Beyond  Standard Model (BSM) Higgs Physics. The main reason for this is that, at these TeV energies, the production of two Higgs bosons proceeds dominantly via the scattering subprocess $WW \to HH$ (usually called $WW$ fusion) where the two $W$ gauge bosons are radiated from the initial colliding electrons and positrons.  The full process of our interest here is then $e^+e^- \to HH \nu \bar \nu$ occurring via $WW$ fusion,  like in the generic \figref{plot-ee}.   
This $WW$ fusion subprocess is in turn highly dominated by the contribution from the longitudinal modes,  $W_LW_L \to HH$,  which are the most sensitive ones to the BSM Higgs couplings at these high energies, specially in the bosonic sector.  One indirect  but simple way to understand this extreme sensitivity is because the $W_L$ modes, by virtue of the Equivalence Theorem (ET),   behave at large energies,   $\sqrt{s}\gg m_W$,  as the Goldstone bosons (GBs) $w$ associated to the spontaneous electroweak symmetry breaking, $SU(2)_L \times U(1)_Y \to U(1)_{\rm em}$,  
and the corresponding GB scattering,  $ww \to HH$,  provides an excellent window to the typical derivative couplings involved in the scalar sector of these BSM theories,   which in turn give enhanced cross sections at the TeV energies. 

Testing the new Higgs couplings involved in the  $WW \to HH$ subprocess is therefore one of our main goals in this work.  For the present analysis,  we will assume that all the particle couplings to the fermions are like the SM ones,  and that the new Higgs physics appears only in the bosonic sector.  We will do this test of BSM Higgs couplings in the bosonic sector,  by means of the two most popular Effective Field Theories (EFTs),  nowadays widely employed in collider physics: the so-called HEFT (Higgs Effective Field Theory) and the SMEFT (Standard Model Effective Field Theory).  The advantage of using EFTs is that they allow for a description of the relevant scattering,  here $WW \to HH$,  in a model independent way.  The information of the anomalous Higgs couplings is encoded in a set of effective operators,  built with the SM fields and with the unique requirement of being invariant under the SM gauge symmetry,  $SU(3) \times SU(2) \times U(1)$.  The coefficients in front of these operators (usually called Wilson coefficients) are generically  unknown and  encode the information of the particular underlying fundamental theory that generates such EFT at low energies,  when the new heavy modes of this theory are integrated out.  It is well known that depending on the kind of dynamics involved in the fundamental theory,  it is more appropriate the use of one EFT or another.  Usually, the SMEFT is more appropriate to describe the low energy behaviour of weakly interacting dynamics,  whereas the HEFT is more appropriate to describe strongly interacting underlying dynamics (for  reviews,  see for instance,  \cite{Brivio:2017vri,Dobado:2019fxe}).  

We will present first the computation of the cross section,  $\sigma(WW \to HH)$,   within the HEFT and then we will compare it with the corresponding cross section within the SMEFT.  We will not use the Equivalence Theorem,  but we will consider instead all the physical gauge boson modes,  longitudinal $W_L$ and transverses $W_T$ in the computation of the scattering amplitudes.   For the HEFT,  since we are interested in the bosonic sector, we will use the Electroweak Chiral Lagrangian (EChL) which contains all the relevant bosonic interactions for the present work.   In the SMEFT,   the effective operators are ordered in terms of their canonical dimension (dim 6, dim 8 etc),  whereas in the HEFT with the  EChL the order of the effective operators is in terms of their chiral dimension ($\chi_{\rm dim}$=2, $\chi_{\rm dim}$=4, etc).  Since, at lowest order in the EChL case (the so-called Leading Order (LO) with effective operators of $\chi_{\rm dim}$=2),   the consequences of  Higgs anomalous couplings at the TeV $e^+e^-$ colliders  have already been studied in the literature, Ref.~\cite{Gonzalez-Lopez:2020lpd},  we will focus here instead in the Next to Leading Order effective operators with $\chi_{\rm dim}$=4.  The comparison with the SMEFT prediction must therefore go beyond its lowest order,  with dim 6 operators,  and include consistently the most relevant dim 8 effective operators.  The interest of this HEFT-SMEFT comparison is that we will be able to determine the correct matching of the two approaches at the level of scattering amplitudes of  $WW \to HH$.  Considering the most relevant coefficients for this scattering in both EFTs,   and  from this matching of amplitudes we will be able to extract  the proper relations among their corresponding Lagrangian coefficients.  This will allow us to fully describe the wanted  BSM Higgs physics in terms of just a few most relevant coefficients with a clear relation among the two HEFT and SMEFT approaches. The second part of this work is the study of the sensitivity to those coefficients at the future TeV $e^+e^-$ colliders.  We will focus in two projected cases,  the International Linear Collider (ILC) \cite{Strube:2016eje, Bambade:2019fyw} with energy $\sqrt{s}=1$ TeV  and luminosity 8 ab$^{-1}$,  and the Compact Linear Collider (CLIC) \cite{Abramowicz:2016zbo, CLICdp:2018cto, Roloff:2019crr} with energy  $\sqrt{s}=3$ TeV and luminosity 5 ab$^{-1}$.  In the final part of this work we will determine the accessible region in these two planned colliders to the most relevant EFT parameters,  by means of the study of the particular final state $b\bar b b \bar b \nu \bar \nu$,  resulting from the decays to quark bottoms of the two Higgs bosons and  leading to a characteristic signal with  four $b$-jets and missing energy. 

The paper is organized as follows: we review the relevant HEFT Lagrangian and present the analytical amplitude for the $WW\to HH$ scattering in \secref{sec:HEFT}. Also in this section, we study the corresponding cross section and identify which operators are the dominant for each polarization state at TeV energy scale. 
In \secref{sec:SMEFT}, we present a similar analysis in the SMEFT context. 
Then, in \secref{sec:matching}, we compare the resulting amplitudes in both approaches and by matching them we obtain the relation among the coefficients in the HEFT and SMEFT. 
Finally, we move to the $e^+e^-$ collider scenario and study the sensitivity to the most relevant EFT coefficients in \secref{sec:sensitivity}.
The conclusions are given in \secref{sec:conclus}.

\section{$WW\to HH$ in HEFT}
\label{sec:HEFT}
In this section we present our study of $WW\to HH$ within the HEFT context.  For this study we select the bosonic sector, containing the GBs, the Higgs field and the EW gauge bosons,  and use the EChL,  which uses a non-linear parametrization of the GB fields and organizes the set of effective operators describing the new Higgs Physics in terms of their chiral dimension.  We will perform this study at the tree level approximation and will consider operators of both types,  the lowest order chiral dimension two,  and of the next to leading order chiral dimension four.  First we present the relevant Lagrangian,  then the relevant scattering amplitude, and then the numerical predictions for the cross sections. 

\subsection{The relevant HEFT Lagrangian}
The relevant HEFT Lagrangian for the present computation is the EChL.  The bosonic fields and building blocks of the EChL are as follows.  The four EW gauge bosons, $W^a_\mu$ ($a=1,2,3$) and $B_\mu$,   that are the interaction eigenstates associated to the $SU(2)_L$ and $U(1)_Y$ symmetries,  respectively,  the three GBs $w^a$ ($a=1,2,3$) associated to the spontaneous breaking $SU(2)_L \times U(1)_Y \to U(1)_{\rm em}$,  and the Higgs boson $H$.  
The GBs are introduced in a non-linear representation, usually via the exponential parametrization,  by means of the unitary matrix $U$:
\be 
U(w^a) = e^{i w^a \tau^a/\vev} \, \, , 
\label{expo}
\ee
where, $\tau^a$, $a=1,2,3$,  are the Pauli matrices and $v=246$ GeV.  
Under an EW chiral transformation of  $SU(2)_L \times SU(2)_R$,  given by $L \in SU(2)_L$ and $R \in SU(2)_R$, the field $U$ transforms linearly as $L U R^\dagger$, whereas the GBs $w^ a$ transform non-linearly.  This peculiarity implies multiple GBs interactions in the HEFT,  not just among themselves but also with the other fields, and it is the main feature of this non-linear EFT.
The $H$ field is, in contrast to the GBs,  a singlet of the EW chiral symmetry and the EW gauge symmetry and, consequently, there are not limitations from symmetry arguments on the implementation of this field and its interactions with itself and with the other fields.  Usually,  in the EChL, the interactions of $H$  are introduced via generic polynomials.

The EW gauge bosons are introduced in the EChL by means of the $SU(2)_L \times U(1)_Y$ gauge prescription,  namely, via the covariant derivative of the $U$ matrix,  and by the $SU(2)_L$ and $U(1)_Y$  field strength tensors,  given by:
\bear
D_\mu U &=& \partial_\mu U + i\hat{W}_\mu U - i U\hat{B}_\mu \,, \nn\\
\hat{W}_{\mu\nu} &=& \partial_\mu \hat{W}_\nu - \partial_\nu \hat{W}_\mu + i  [\hat{W}_\mu,\hat{W}_\nu ] \,, 
\quad \hat{B}_{\mu\nu} = \partial_\mu \hat{B}_\nu -\partial_\nu \hat{B}_\mu \,,  
\eear
where $\hat{W}_\mu = g W^a_\mu \tau^a/2$, and  $\hat{B}_\mu = \gY B_\mu \tau^3/2$.  For the construction of the EChL and 
in addition to these basic building blocks,  it is also customary to use the following objects:
\bear
 \mV_\mu&=&(D_\mu U)U^\dagger   \,, 
\quad {\cal D}_\mu O=\partial_\mu O+i[\hat{W}_\mu,O]\, .
\eear
The physical gauge fields are then given,  as usual,  by:
\be
W_{\mu}^\pm = \frac{1}{\sqrt{2}}(W_{\mu}^1 \mp i W_{\mu}^2) \,,\quad
Z_{\mu} = c_W W_{\mu}^3 - s_W B_{\mu} \,,\quad
A_{\mu} = s_W W_{\mu}^3 + c_W B_{\mu} \,,
\label{eq-gaugetophys}
\ee
where we use the short notation $s_W=\sin \theta_W$ and $c_W=\cos \theta_W$,  with $\theta_W$ the weak angle.

We consider here only the relevant effective operators for the scattering of our interest,  $WW \to HH$,  and restrict ourselves to the subset that is invariant under the custodial symmetry,  an approximation which is very reasonable for the study of this scattering process at TeV energies.  The operators selected in the EChL are  organized as usual by their chiral dimension into two terms:  ${\cal L}_2$,  with chiral dimension two and  ${\cal L}_4$
with chiral dimension four.  In momentum space, a $\chi_{\rm dim}$=2 contribution means ${\cal O}(p^2)$ whereas a $\chi_{\rm dim}$=4 contribution means ${\cal O}(p^4)$.  For this chiral counting,  we consider as usual that all involved masses count equally as momentum,  namely,  with chiral dimension one.  Consequently,  $\partial_\mu \,,\,\mw \,,\,\mz \,,\,\mh \,,\,g\vev \,,\,\gY\vev \,,\lambda v\, \sim \mO(p) \,$.  Thus, 
 the relevant  EChL, that is  $SU(2)_L \times U(1)_Y$ gauge (and custodial)  invariant, and that is valid for NLO tree level calculations which include $\chi_{\rm dim}$=2 and $\chi_{\rm dim}$=4 operators, is summarized  by:
\be
{\cal L}_{\rm EChL}=\mL_2+ \mL_4 \, 
\label{EChL}
\ee
 where the relevant chiral dimension two  Lagrangian for $WW \to HH$ is
\bear
\mL_2 &=& \frac{v^2}{4}\left(1+2a\frac{H}{v}+b\frac{H^2}{\vev^2}\right){\rm Tr}\Big[ 
 D_\mu U^\dagger D^\mu U \Big]+\frac{1}{2}\partial_\mu H\partial^\mu H-V(H)  \nn\\
&&-\frac{1}{2g^2} {\rm Tr}\Big[ \hat{W}_{\mu\nu}\hat{W}^{\mu\nu}\Big]
-\frac{1}{2\gYc}{\rm Tr}\Big[ 
\hat{B}_{\mu\nu}\hat{B}^{\mu\nu}\Big]  +\mL_{GF} +\mL_{FP} \,, 
\label{eq-L2}
\eear
and  the relevant chiral dimension four  Lagrangian for  $WW \to HH$ is
\bear
{\mL}_{4}&=& -a_{dd{\cal V}{\cal V}1} \frac{\partial^\mu H\,\partial^\nu H}{v^2} {\rm Tr}\Big[ {\cal V}_\mu {\cal V}_\nu \Big] 
-a_{dd{\cal V}{\cal V}2} \frac{\partial^\mu H\,\partial_\mu H}{v^2} {\rm Tr}\Big[ {\cal V}^\nu {\cal V}_\nu \Big] +a_{11} {\rm Tr}\Big[{\cal D}_\mu {\cal V}^\mu {\cal D}_\nu {\cal V}^\nu\Big]  \nn\\
&& -
\frac{\mh^2}{4}\left(2a_{H\mV\mV}\frac{H}{v}+a_{HH\mV\mV}\frac{H^2}{\vev^2} \right){\rm Tr}\Big[ {\cal V}^\mu {\cal V}_\mu \Big]  \nn\\
&& - \left(a_{HWW} \frac{H}{v} +a_{HHWW} \frac{H^2}{v^2}\right) {\rm Tr}\Big[\hat{W}_{\mu\nu} \hat{W}^{\mu\nu}\Big] +i\left(a_{d2} +a_{Hd2}\frac{H}{v}\right)\frac{\partial^\nu H}{v} {\rm Tr}\Big[ \hat{W}_{\mu\nu} {\cal V}^\mu\Big]  \nn\\
&&  +\left(a_{\Box\mV\mV} +a_{H\Box\mV\mV}\frac{H}{\vev}\right)\frac{\Box H}{\vev} {\rm Tr}\Big[{\cal V}_\mu {\cal V}^\mu\Big] +a_{d3}\frac{\partial ^\nu H}{\vev} {\rm Tr}\Big[{\cal V}_\nu {\cal D}_\mu {\cal V}^\mu \Big]   \nn\\
&&+\left(a_{\Box\Box} +a_{H\Box\Box}\frac{H}{\vev}\right)\frac{\Box H\,\Box H}{\vev^2} +a_{dd\Box}\frac{\partial^\mu H\,\partial_\mu H\,\Box H}{\vev^3} + a_{Hdd}\frac{\mh^2}{\vev^2}\frac{H}{\vev}\partial^\mu H\,\partial_\mu H 
\label{eq-L4-without-eoms}
\eear
In the  Lagrangian with $\chi_{\rm dim}$= 2,  in \eqref{eq-L2},  $ \mL_{GF}$ and  $\mL_{FP}$ are the gauge-fixing  Lagrangian and Fadeev-Popov  Lagrangian,  respectively,  and $V(H)$ is the Higgs potential,  which we take here as
\be
V(H) = \frac{1}{2}\mh^2 H^2 +\kappa_3\lambda\vev H^3+\kappa_4\frac{\lambda}{4}H^4 \,,
\label{EChLpotential}
\ee
with $m_H^2= 2 \lambda v^2$.  Concretely,  for the present computation of the $WW \to HH$ scattering we will set the Feynman-'t Hooft gauge with gauge fixing parameter $\xi=1$.  The specific formulas for $\mL_{GF}$  and $\mL_{FP}$ in generic  covariant $R_\xi$ gauges, within the EChL context,  can be found in~\cite{Herrero:2021iqt}. 

The reference values for the coefficients in the EChL  to reach the SM predictions are: $a=b=\kappa_3=\kappa_4=1$ in  $\mL_2$,  and $a_i=0$ for all the coefficients in  $\mL_4$.   This means that
the new physics BSM is encoded in the chiral coefficients $a$, $b$, $\kappa_3$ and $\kappa_4$ of $\mL_2$ when they are different from one,  and in the non-vanishing values of the $a_i$ coefficients of $\mL_4$.  

The previous Lagrangian with $\chi_{\rm dim}$=4 in \eqref{eq-L4-without-eoms} can be further reduced by the use of the equations of motion (EOMs) if these operators  are to be used in a tree level computation of a scattering amplitude where the external legs are on-shell,  like the one we are interested in here.  
Then, one can rewrite the operators including the $\Box H$ or ${\cal D}_\mu \mV^\mu$ pieces in terms of other operators in ${\mL}_4$ by using the following equations,  as in~\cite{Brivio:2013pma}:
\bear
\Box H &=& -\frac{\delta V(H)}{\delta H} -\frac{\vev^2}{4}\frac{\mF(H)}{\delta H}{\rm Tr}\Big[ {\cal V}^\mu {\cal V}_\mu \Big] \, ,  \nn\\
{\rm Tr}\Big[ \tau^j{\cal D}_\mu \mV^\mu \Big] \mF(H) &=& -{\rm Tr}\Big[ \tau^j \mV^\mu \Big] \partial_\mu\mF(H) \,\, ,
\label{general-eoms}
\eear
where
\bear
{\mF(H)}&=&\left(1+2a\frac{H}{v}+b\frac{H^2}{\vev^2}\right) \,.
\label{gothicF}
\eear

In particular, for the present scattering $WW \to HH$ with external $W^\pm$ and $H$ on-shell states, one can use the following EOMs, where we have kept in \eqref{general-eoms} just the terms that provide a maximum of two $H$ or two $W$ gauge bosons in the operator: 
\bear
\Box H &=& -\mh^2 H -\frac{3}{2}\kappa_3\mh^2\frac{H^2}{\vev} -\frac{a}{2}\vev{\rm Tr}\Big[ {\cal V}^\mu {\cal V}_\mu \Big] -\frac{b}{2}H{\rm Tr}\Big[ {\cal V}^\mu {\cal V}_\mu \Big] \, ,  \nn\\
{\rm Tr}\Big[ \tau^j{\cal D}_\mu \mV^\mu \Big] &=& -{\rm Tr}\Big[ \tau^j \mV^\mu \Big] \frac{2a}{\vev}\partial_\mu H \, .
\label{our-eoms}
\eear
   
Thus,  it is convenient to use the simplified Lagrangian that is obtained after the use of these EOMs in \eqref{our-eoms} which is written in terms of a reduced set of couplings. 
In particular,  the operators of $a_{11}$ and $a_{d3}$ can be written in terms of the operator of $a_{dd\mV\mV 1}$; the operator of $a_{dd\Box}$ in terms of the operators of $a_{dd\mV\mV 2}$ and $a_{Hdd}$; and the operators of
$a_{\Box\mV\mV}$, $a_{H\Box\mV\mV}$, $a_{\Box\Box}$, and $a_{H\Box\Box}$ in terms of the operators of $a_{H\mV\mV}$ and $a_{HH\mV\mV}$ (and also with other operators which do no enter in this observable). Thus,  after the use of the EOMs there is just the reduced basis of operators with the corresponding  combinations of coefficients which can be renamed again, to simplify,  as in the starting Lagrangian.  For instance,  the combination of coefficients entering in the first operator of \eqref{eq-L4-without-eoms} after the use of the EOMs is $(-a_{dd{\cal V}{\cal V}1}-4 a^2 a_{11}+ 2 a a_{d3})$ which we rename as $(-a_{dd{\cal V}{\cal V}1})$,  and that for the second operator is $(-a_{dd{\cal V}{\cal V}2}+(a/2) a_{dd\Box})$ which we rename as  $(-a_{dd{\cal V}{\cal V}2})$. 

Finally, for the present computation of $WW\to HH$, this reduced version of ${\mL}_4$ can be written as follows:
\bear
\hspace{-10mm}{\mL}_{4}^{\rm +EOMs}&=& -a_{dd{\cal V}{\cal V}1} \frac{\partial^\mu H\,\partial^\nu H}{v^2} {\rm Tr}\Big[ {\cal V}_\mu {\cal V}_\nu \Big] 
-a_{dd{\cal V} {\cal V}2} \frac{\partial^\mu H\,\partial_\mu H}{v^2} {\rm Tr}\Big[ {\cal V}^\nu {\cal V}_\nu \Big]  \nn\\
&& -\frac{\mh^2}{4}\left(2a_{H\mV\mV}\frac{H}{v}+a_{HH\mV\mV}\frac{H^2}{\vev^2} \right){\rm Tr}\Big[ {\cal V}^\mu {\cal V}_\mu \Big] +a_{Hdd}\frac{\mh^2}{\vev^2}\frac{H}{\vev}\partial^\mu H\,\partial_\mu H  \nn\\
&& - \left(a_{HWW} \frac{H}{v} +a_{HHWW} \frac{H^2}{v^2}\right) {\rm Tr}\Big[\hat{W}_{\mu\nu} \hat{W}^{\mu\nu}\Big] +i\left(a_{d2} +a_{Hd2}\frac{H}{v}\right)\frac{\partial^\nu H}{v} {\rm Tr}\Big[ \hat{W}_{\mu\nu} {\cal V}^\mu\Big]
\label{eq-L4-with-eoms}
\eear
In summary,  our starting EChL contains the following relevant coefficients for $WW \to HH$: 3 coefficients  in $\mL_2$,  $a$,  $b$ and 
$\kappa_3$ ($\kappa_4$ does not enter in this process) and 9 coefficients  in $\mL_4$,  $a_{dd\mV \mV 1}$, $a_{dd\mV\mV 2}$,  $a_{d2}$,  $a_{Hd2}$,  $a_{Hdd}$,  $a_{HWW}$,  $a_{HHWW}$,  $a_{H \mV \mV}$ and $a_{HH\mV \mV}$. 
Notice, that other scattering processes different than $WW \to HH$ would require a different set of reduced operators in $\mL_4$.  For a complete list of effective operators in the HEFT,  see for instance~\cite{Brivio:2013pma}.  Notice also that we have used here a different notation than in that reference.  The relation among the two sets of coefficients can be summarized by: 
$a_{dd\mV\mV 1}\leftrightarrow c_8$, $a_{dd\mV\mV 2}\leftrightarrow c_{20}$, $a_{11}\leftrightarrow c_9$, $a_{HWW}\leftrightarrow a_W$, $a_{HHWW}\leftrightarrow b_W$, $a_{d2}\leftrightarrow c_5$, $a_{Hd2}\leftrightarrow a_5$, $a_{\Box\mV\mV}\leftrightarrow c_7$, $a_{H\Box\mV\mV}\leftrightarrow a_7$, $a_{d3}\leftrightarrow c_{10}$, $a_{Hd3}\leftrightarrow a_{10}$, $a_{\Box\Box}\leftrightarrow c_{\Box H}$, $a_{H\Box\Box}\leftrightarrow a_{\Box H}$, $a_{dd\Box}\leftrightarrow c_{\Delta H}$, $a_{H\mV\mV}\leftrightarrow a_{C}$ and $a_{HH\mV\mV}\leftrightarrow b_{C}$.

\subsection{Scattering amplitude in HEFT}
\label{sec:ampHEFT}
Here and in the following of this work,  the notation for the momenta assignments and Lorentz indexes involved in the scattering of our interest $WW \to HH$  is set as follows:
\be
W^+_\mu(p_+)\,W^-_\nu(p_-) \to H(k_1)\,H(k_2)\,,
\label{ourscattering}
\ee
where $p_\pm$ and $k_{1,2}$ (with $p_+ +p_- =k_1 +k_2$) are the incoming and outcoming momenta of the bosons. The $W^{\pm}$ polarization vectors are $\epsilon_\pm$,  respectively.

For the computation of the amplitude from  the EChL, we work at the tree level, set the Feynman-'t Hooft gauge (i.e. with massive GBs) and write the result in terms of the corresponding contributions from the different channels: the S-channel, the T-channel, the U-channel, and the contact term.  Notice that in this section we use capital letters for the $s$, $t$ and $u$ Mandelstam variables.  
Thus, the tree level amplitude within the HEFT at NLO is given by:
\bear
 \amp(WW \to HH)\vert_{\rm HEFT}  &=&  \amp^{(2)} +  \amp^{(4)}
 \label{ampWWtoHHtree}
\eear
where the chiral-dim 2 and chiral-dim 4 contributions are given,  respectively, by:
\bear
 \amp^{(2)}&=&\amp^{(2)}\vert_S+\amp^{(2)}\vert_T+\amp^{(2)}\vert_U+\amp^{(2)}\vert_C \nn\\
  \amp^{(4)}&=&\amp^{(4)}\vert_S+\amp^{(4)}\vert_T+\amp^{(4)}\vert_U+\amp^{(4)}\vert_C
 \label{ampWWtoHH2and4}
 \eear
 with the corresponding contributions from the various channels  given by:

\bear
\amp^{(2)}\vert_S &=& \frac{g^2}{2}3a\kappa_3\frac{\mh^2}{S-\mh^2}\epsilon_{+}\cdot\epsilon_{-}  \nn\\
\amp^{(2)}\vert_T &=& g^2a^2\frac{\mw^2\epsilon_{+}\cdot\epsilon_{-} +\epsilon_{+}\cdot k_1\,\epsilon_{-}\cdot k_2}{T-\mw^2}  \nn\\
\amp^{(2)}\vert_U &=& g^2a^2\frac{\mw^2\epsilon_{+}\cdot\epsilon_{-} +\epsilon_{+}\cdot k_2\,\epsilon_{-}\cdot k_1}{U-\mw^2}  \nn\\
\amp^{(2)}\vert_C &=& \frac{g^2}{2}b\,\epsilon_{+}\cdot\epsilon_{-}  \nn\\
\amp^{(4)}\vert_S &=& \frac{g^2}{2\vev^2}\frac{1}{S-\mh^2}\left(3\kappa_3a_{d2}\mh^2(S\epsilon_+\cdot\epsilon_- -2\epsilon_+\cdot p_-\,\epsilon_-\cdot p_+)  \right.  \nn\\
&&\left. \hspace{3mm}+6\kappa_3a_{HWW}\mh^2((S-2\mw^2)\epsilon_+\cdot\epsilon_- -2\epsilon_+\cdot p_-\,\epsilon_-\cdot p_+)  \right.  \nn\\
&&\left. \hspace{3mm}-(3\kappa_3a_{H\mV\mV}\mh^4+aa_{Hdd}\mh^2(S+2\mh^2))\epsilon_+\cdot\epsilon_- \right)  \nn\\
\amp^{(4)}\vert_{T} &=& \frac{g^2}{2\vev^2}\frac{a}{T-\mw^2} \left(a_{d2}(4\mw^2\mh^2\epsilon_+\cdot\epsilon_- +2(T+3\mw^2-\mh^2)\epsilon_+\cdot k_1 \epsilon_-\cdot k_2  \right.  \nn\\
&& \left. \hspace{30mm}-4\mw^2(\epsilon_+\cdot k_1\epsilon_-\cdot p_+ +\epsilon_+\cdot p_-\epsilon_-\cdot k_2 ))  \right.  \nn\\
&& \left. \hspace{12mm}-8a_{HWW}\mw^2((T+\mw^2-\mh^2)\epsilon_+\cdot\epsilon_- +\epsilon_+\cdot k_1 \, \epsilon_-\cdot p_+ +\epsilon_+\cdot p_- \, \epsilon_-\cdot k_2)  \right.  \nn\\
&& \left. \hspace{12mm}-4a_{H\mV\mV}\mh^2(\mw^2\epsilon_{+}\cdot\epsilon_{-} +\epsilon_{+}\cdot k_1\,\epsilon_{-}\cdot k_2)  \right)  \nn\\  
\amp^{(4)}\vert_{U} &=& \amp^{(4)}\vert_{T} \,\,\,{\rm with}\,\,\,T\to U \,\,\,{\rm and}\,\,\, k_1\leftrightarrow k_2  \nn\\
\amp^{(4)}\vert_C &=& \frac{g^2}{2\vev^2}\left(-2 a_{dd{\cal V}{\cal V}1} (\epsilon_+\cdot k_2 \, \epsilon_-\cdot k_1+\epsilon_+\cdot k_1 \, \epsilon_-\cdot k_2)  \right.\nn\\
&&\left. \hspace{3mm}+(-2a_{dd{\cal V}{\cal V}2}(S-2\mh^2) +4a_{HHWW}(S-2\mw^2) +a_{Hd2}S -a_{HH\mV\mV}\mh^2)\epsilon_+\cdot\epsilon_-  \right.\nn\\
&&\left. \hspace{3mm}-2 (a_{Hd2}+4 a_{HHWW}) \epsilon_+\cdot p_- \, \epsilon_-\cdot p_+ \right)
\label{ampWWtoHH-HEFT}
\eear
Notice that the SM prediction is reached from the previous HEFT result by taking  $a=b=\kappa_3=1$ ($\kappa_4$ does not enter in this scattering  amplitude) and all the remaining EChL  coefficients in ${\mL}_{4}$ set to zero, $ a_i=0$.  
We include the SM amplitude below for comparison,  where we also use the Feynman-'t Hooft gauge and $m_H^2=2 \lambda v^2$,  $m_W^2=g^2v^2/4$,  
\bear
 \amp(WW \to HH) \vert_{\rm SM} &=& \frac{g^2}{2} 3 \frac{\mh^2}{S-\mh^2}\epsilon_{+}\cdot\epsilon_{-}  \nn\\
 && +g^2 \frac{\mw^2\epsilon_{+}\cdot\epsilon_{-} +\epsilon_{+}\cdot k_1\,\epsilon_{-}\cdot k_2}{T-\mw^2}  \nn\\
&& +g^2 \frac{\mw^2\epsilon_{+}\cdot\epsilon_{-} +\epsilon_{+}\cdot k_2\,\epsilon_{-}\cdot k_1}{U-\mw^2}  \nn\\
&& +\frac{g^2}{2} \,\epsilon_{+}\cdot\epsilon_{-}  
\label{ampWWtoHH-SM}
\eear
Some comments are in order about the previous analytical results for the scattering amplitude  $\amp(WW \to HH)\vert_{\rm HEFT}$.  First of all, notice that it is written  in terms of the polarization vectors of the initial $W^+$ and $W^-$,   given by  $\epsilon_+(p_+)$ and $\epsilon_-(p_-)$,  respectively.  Therefore, the particular physical helicity amplitudes for the polarized gauge bosons,  longitudinal $W_L$ and transverse $W_T$,  can be obtained from these expressions above, just by replacing the corresponding polarization vectors for these $L$ and $T$ modes.   The expressions of $ \amp^{(2)}$ above have been found previously in
the literature,  Ref.~\cite{Gonzalez-Lopez:2020lpd}, and we have checked the agreement with those results.  The expressions of $ \amp^{(4)}$ above are new. 
Other previous computations in the literature,  contain some simplifications.  In~\cite{Delgado:2013hxa} this amplitude is computed using the Equivalence Theorem,  therefore,  it is computed with GBs $w$ in the external legs. The corresponding amplitude $\amp^{(4)}$ of  the GB scattering  $\amp(ww \to HH)$ is given in terms of two coefficients,  called $\eta$ and $\delta$ in~\cite{Delgado:2013hxa},   that correspond to the two scalar operators of $\chi_{\rm dim}$=4  that are the relevant ones in that case.  Concretely,  
\bear
{\mL}_{4}^{\rm scalar}&=& \eta  \frac{\partial^\mu H\,\partial^\nu H}{v^2} {\rm Tr}\Big[ \partial_\mu U^\dagger \partial_\nu U \Big] + \delta \frac{\partial^\mu H\,\partial_\mu H}{v^2} {\rm Tr}\Big[ \partial^\nu U^\dagger \partial_\nu U \Big] +\dots
\label{L4-ET}
\eear
Notice that this scalar part is contained in our  EChL Lagrangian.  Concretely,  this  is inside the first two terms of \eqref{eq-L4-with-eoms} which can be rewritten as,
\bear
\mL_4&=& a_{dd\mV \mV 1} \frac{\partial^\mu H\,\partial^\nu H}{v^2} {\rm Tr}\Big[ D_\mu U^\dagger {D}_\nu U \Big]
 +a_{dd\mV \mV 2} \frac{\partial^\mu H\,\partial_\mu H}{v^2} {\rm Tr}\Big[ {D}^\nu U^\dagger {D}_\nu U \Big]+\dots
 \label{addV1+addV2}
 \eear
Thus,  the relation of the two above Lagrangians in \eqrefs{L4-ET}{addV1+addV2} can be simply obtained by replacing  $a_{dd\mV \mV 1}$ by $\eta$ and $a_{dd\mV \mV 2}$ by $\delta$. 

The tree level analytical computation of $ \amp^{(4)}$  in~\cite{Asiain:2021lch} is performed with physical external gauge bosons and with a reduced set of effective operators in the EChL. Their  analytical results for the various channels in the $WW \to HH$ scattering amplitude are given in the Landau gauge,  i.e.  with massless GBs.  The reduced set of effective operators and  $\chi_{\rm dim}$=4 coefficients, named $\eta$, $\delta$ and $\chi$  that are involved in the results of this reference correspond to our $a_{dd\mV \mV 1}$,  $a_{dd\mV \mV 2}$,  
$a_{d2}$ and $a_{Hd2}$.  We have checked the agreement of the contributions to the amplitude for this subset of operators with those results in~\cite{Asiain:2021lch} by doing the following replacement,  
$a_{dd\mV \mV 1}$ by $\eta$,  $a_{dd\mV \mV 2}$ by 
$\delta$,  $a_{d2}$ by $b_1 \chi$ and $a_{Hd2}$ by $2 b_2 \chi$.  The other contributions in the amplitude from the remaining $a_i$'s are not included in~\cite{Asiain:2021lch}.

In summary,  in this section we have determined the scattering amplitude for all the generic polarization channels,  in  $\amp(W_XW_Y  \to HH) \vert_{\rm HEFT}$ with $XY=LL, TT, LT, TL$,  in terms of 3 coefficients of $\mL_2$: $a$, $b$ and 
$\kappa_3$ and of 9 coefficients of $\mL_4$: 
$a_{dd\mV \mV 1}$, $a_{dd\mV\mV 2}$,  $a_{d2}$,  $a_{Hd2}$,  $a_{Hdd}$,  $a_{HWW}$,  $a_{HHWW}$,  $a_{H \mV \mV}$ and $a_{HH\mV \mV}$.  In the next section we will determine which coefficients among these ones are the most relevant coefficients to describe the BSM Higgs physics at the TeV colliders.
\begin{figure}[t!]
\begin{center}
\includegraphics[scale=0.4]{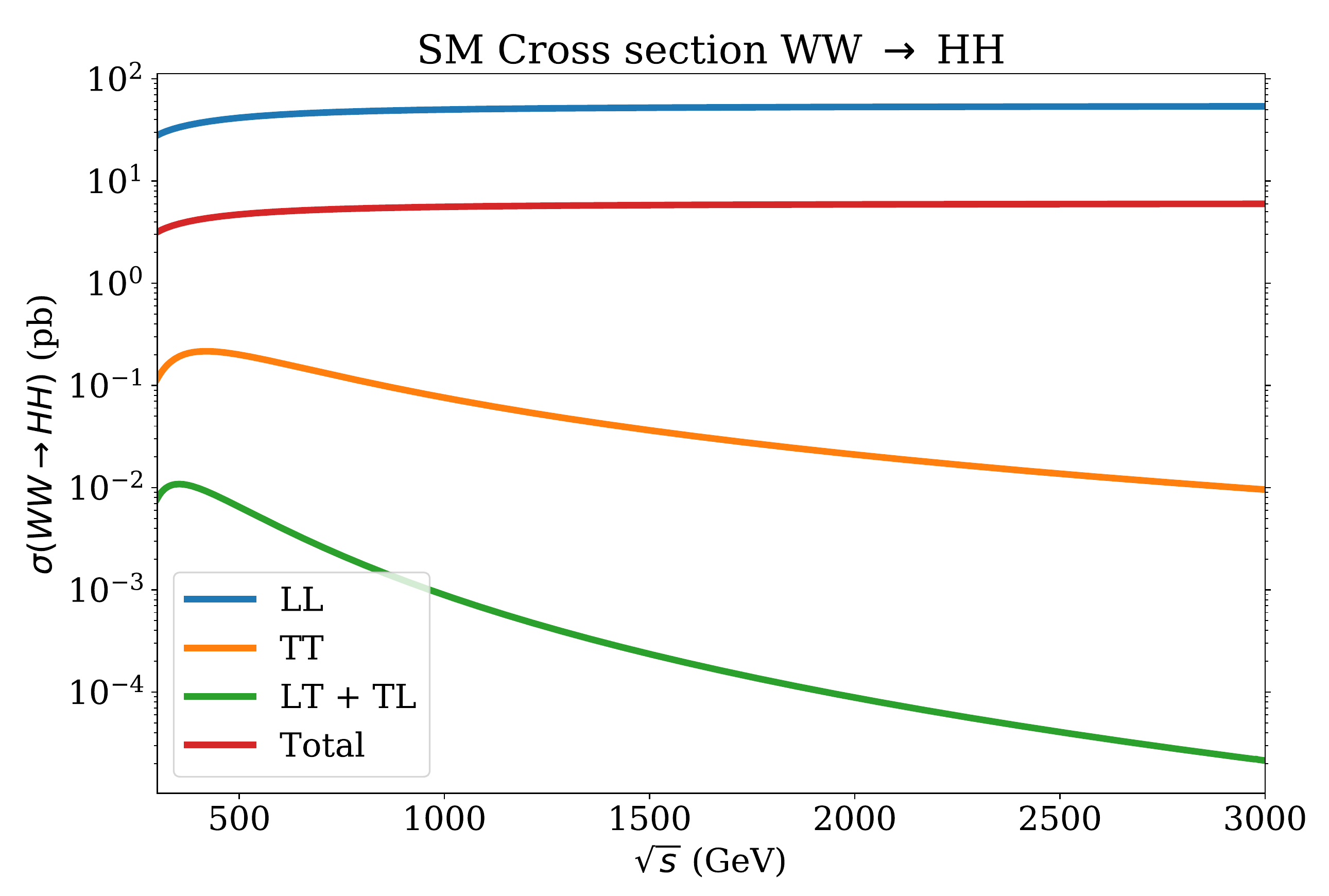} 
\caption{Cross section prediction in the SM for the polarized channels $W_XW_Y \to HH$  as a function of the 
$WW$  center of mass energy $\sqrt{s}$.  The lines are for $XY=LL$ (blue), $XY=TT$ (orange) and $XY=LT+TL$ (green).  The total (unpolarized) SM cross section is also shown (red line).}
\label{SM-xsection-pol}
\end{center}
\end{figure}

\subsection{Cross section results  in HEFT}
\label{sec:xsHEFT}
In this subsection we present the numerical results for the cross section of  $W^+W^- \to HH$.  Since the numerical analysis of the effects from the 3 coefficients in $\mL_2$, $a$, $b$ and $\kappa_3$,  has already been done in the literature~\cite{Gonzalez-Lopez:2020lpd}, we will focus here in the numerical analysis of the effects from the 9 coefficients in  $\mL_4$,  and set $a=b=\kappa_3= 1$.  In principle, all the 9 coefficients contribute to the total (unpolarized) cross section $\sigma(W^+W^- \to HH)$.  However,  in order to understand which are the most relevant coefficients among these 9,  it is very illustrative to compute first the cross section for the polarized modes,  i.e,   $\sigma_{LL}=\sigma(W^+_LW^-_L \to HH)$,   $\sigma_{TT}=\sigma(W^+_TW^-_T\to HH)$ and $\sigma_{LT+TL}=\sigma(W^+_LW^-_T \to HH)+\sigma(W^+_TW^-_L \to HH)$, where the average over the initial helicities is understood (with helicities: 0 for $L$ and  $\pm 1$ for $T$).  In the case of the SM, it is a well known result  the clear dominance of the 
$\sigma_{LL}$ over $\sigma_{TT}$ and  $\sigma_{LT+TL}$ at the TeV energies.  

This is shown in \figref{SM-xsection-pol} where we plot the SM cross section as a function of the $WW$ center-of-mass energy,  separating the various polarization channels,  $LL$, $TT$ and $LT+TL$, and also the total (unpolarized) cross section.  
In fact,  we see that the two lines for the total (red) and for $LL$ (blue)  coincide in the full energy range studied (up the obvious factor 1/9 in the unpolarized cross section due to the average over the initial helicities). Thus,  the total cross section in the SM  is very well approximated at the TeV energies by $\sigma_{LL}$.  Besides,  the SM result,  both for $LL$ and for the total,  shows the well known behaviour with energy,  being flat with $\sqrt{s}$ above around 500 GeV and reaching a constant value,  that for the $LL$ channel is $\sigma_{LL}^{\rm SM} \approx$ 53 pb.  We will see here that this dominance of $\sigma_{LL}$ also happens in the EFT case,  and the size of the total cross section and its behaviour with energy  can be fully understood in terms of the polarized  $\sigma_{LL}$. 
\begin{figure}[t!]
\begin{center}
\hspace{-3mm}\includegraphics[scale=0.46]{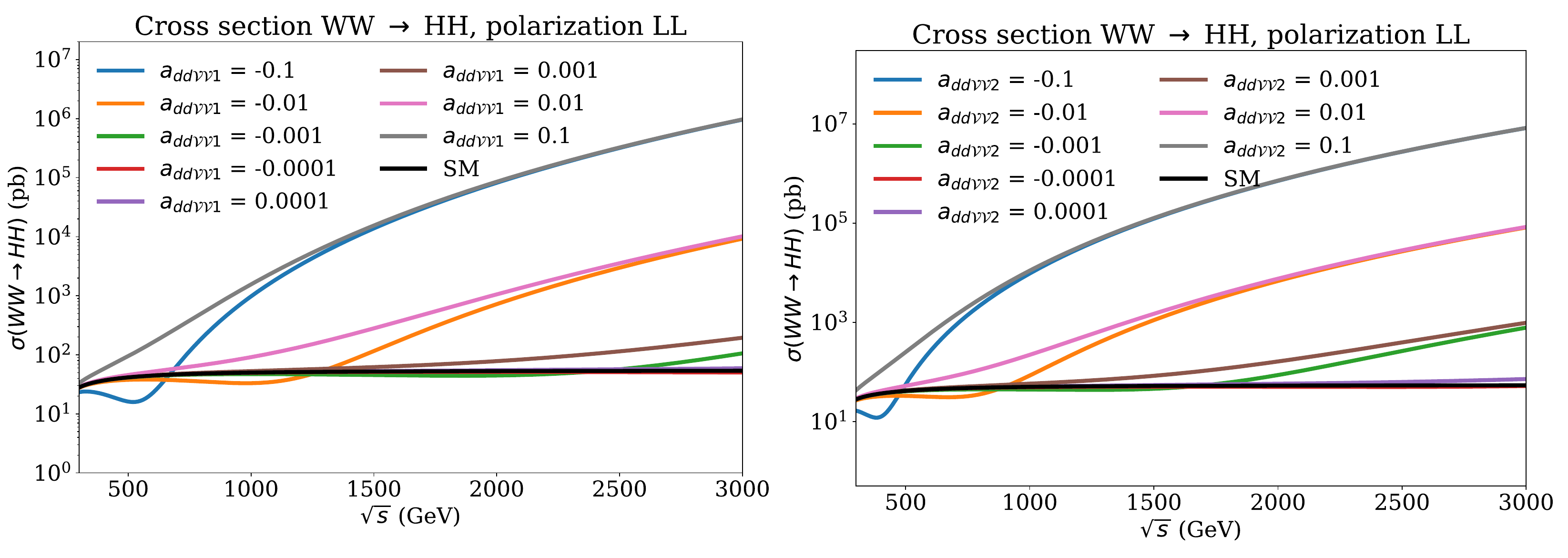} 
\caption{Cross section prediction in the HEFT at subprocess level corresponding to the polarization state $LL$ for different parameter values of $a_{dd\mV\mV1}$ (left) and $a_{dd\mV\mV2}$ (right). The SM prediction (black) is shown for comparison and corresponds to vanishing EChL coefficients.}
\label{LL1}
\end{center}
\end{figure}
\begin{figure}[h!]
\begin{center}
\hspace{-4mm}\includegraphics[scale=0.46]{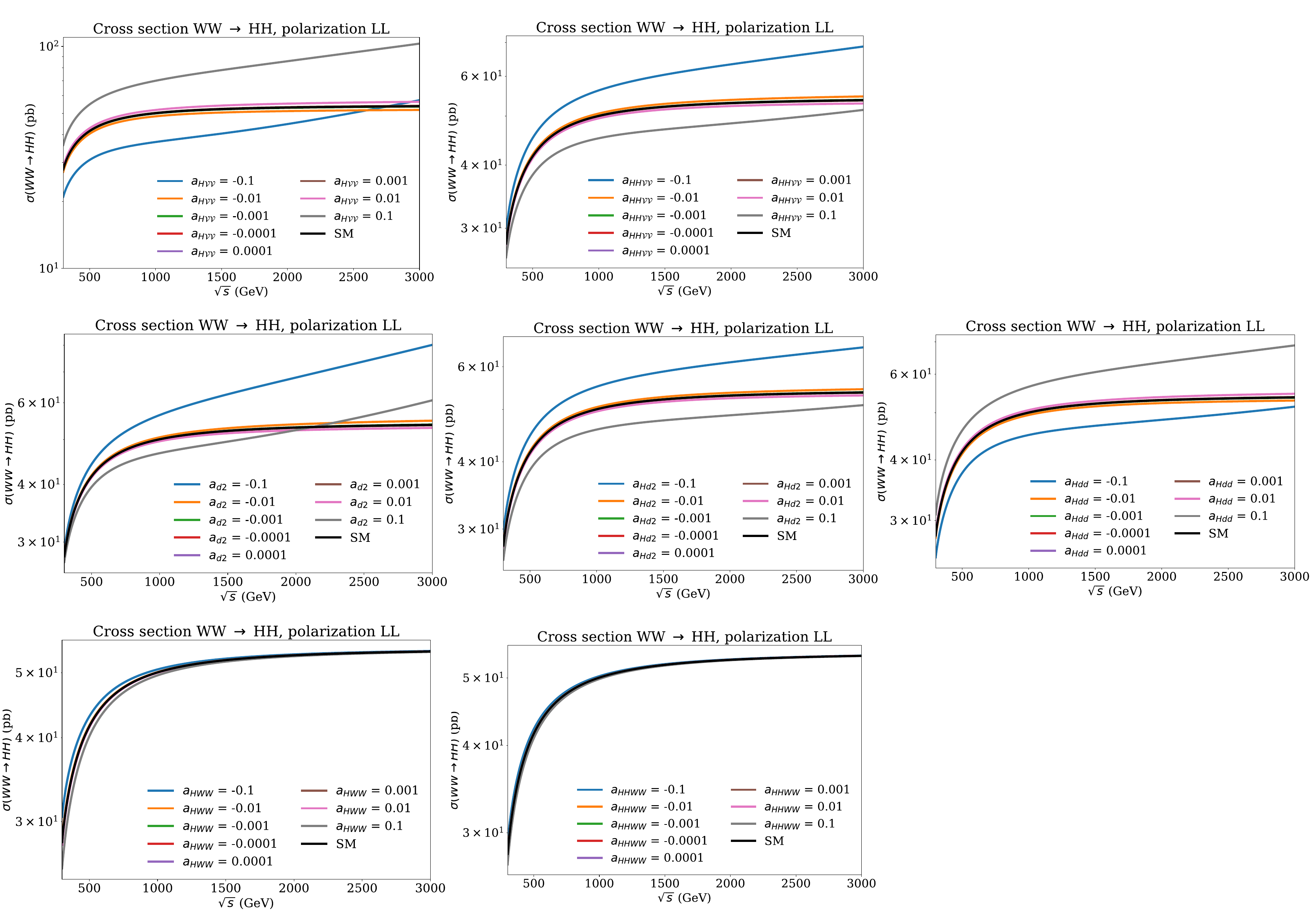} 
\caption{Cross section prediction in the HEFT at subprocess level corresponding to the polarization state $LL$ for different parameter values of $a_{H\mV\mV}$, $a_{HH\mV\mV}$, $a_{d2}$, $a_{Hd2}$, $a_{Hdd}$, $a_{HWW}$ and $a_{HHWW}$,  displayed from left to right and from upper to lower panels,  respectively. The SM prediction (black) is shown for comparison and corresponds to vanishing EChL coefficients.}
\label{LL2}
\end{center}
\end{figure}

The BSM results for $\sigma_{LL}$,  $\sigma_{TT}$ and $\sigma_{LT+TL}$ as a function of the $WW$ center of mass energy $\sqrt{s}$  are shown in \figrefs{LL1}{LT2}.  In each plot we explore the effect of each coefficient $a_i$ separately,  setting the others to zero values.  We explore the cross sections for the following numerical values for the  non-vanishing coefficient:  $\pm 0.1$,  $\pm 0.01$, $\pm 0.001$ and $\pm 0.0001$.  Consequently,  there are 9 plots for each polarization case.  In all plots of these figures the corresponding predictions for the SM case are also included for comparison. 

We start with the analysis of the results for $\sigma_{LL}$ in \figrefs{LL1}{LL2}.  We have selected first  in \figref{LL1} the two most prominent results, meaning the ones with the largest cross sections.  The two coefficients in these plots  of \figref{LL1},   $a_{dd \mV \mV 1}$,  $a_{dd\mV \mV2}$,  are therefore  the most relevant ones,  since for a given assumed numerical value for these two $a_i$'s  they provide sizeable cross sections at the TeV region which are clearly above (by orders of magnitude) the corresponding predictions from the other $a_i$ coefficients having the same assumed value.   Furthermore,  the size of the HEFT cross section for these two coefficients grow  faster with the process energy  than in the other cases.  In particular,  these can be several orders of magnitude above the SM prediction for the cases $a_i=\pm 0.1$ and $a_i=\pm 0.01$.  
The predictions for the other 7 coefficients, less relevant than the two previous ones,  are presented in \figref{LL2}.  It is clear from this figure also the hierarchy in the relevance of the various coefficients, being the ones in the first row more relevant than those on the second row and these in turn more relevant than those in the last row.  

In order to understand the origin of this hierarchy in the relevance of the various coefficients,  we have performed an expansion of the amplitude
$\amp^{(4)}$ in  powers of $s$ (here $s$ denotes the total center-of-mass energy squared,  which was named in the previous section with capital letter as $S$).  For this expansion we have also taken into account the different behaviour with energy of the corresponding polarization vectors.  Then,  for the $LL$  case we have found the following behaviour for the highest ${\cal O}(s^n)$ terms in $\amp^{(4)}(W_LW_L \to HH)$:
\begin{itemize}
\item[1)] The contributions from  $a_{dd \mV \mV 1}$ and $a_{dd\mV \mV2}$ grow as $\sim s^2$
\item[2)] The contributions from  $a_{d2}$,  $a_{Hd2}$,  $a_{Hdd}$, $a_{H \mV \mV}$ and $a_{HH\mV \mV}$ grow as $\sim s^1$
\item[3)] The contributions from $a_{HWW}$ and $a_{HHWW}$ go as $\sim s^0$
\end{itemize}
And this explains the behaviour of the cross sections in \figrefs{LL1}{LL2}.  
\begin{figure}[h!]
\begin{center}
\hspace{-3mm}\includegraphics[scale=0.46]{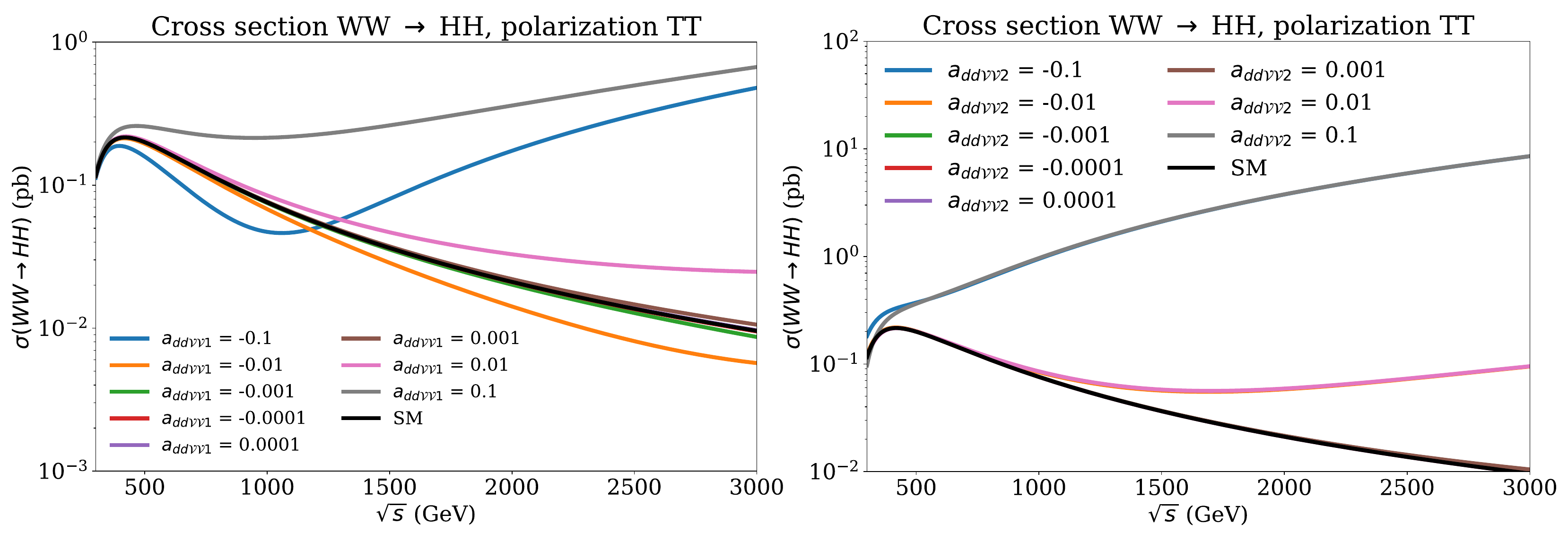} 
\caption{Cross section prediction in the HEFT at subprocess level corresponding to the polarization state $TT$ for different parameter values of $a_{dd\mV\mV1}$ (left) and $a_{dd\mV\mV2}$ (right). The SM prediction (black) is shown for comparison and corresponds to vanishing EChL coefficients.}
\label{TT1}
\end{center}
\end{figure}
 
The results for $\sigma_{TT}$  are shown in \figrefs{TT1}{TT2} where we have set the same order in the presentation of the plots as in the previous $LL$ case,  to make the comparison clear.  We see in \figref{TT1} that the coefficients $a_{dd \mV \mV 1}$ and $a_{dd\mV \mV2}$ are also relevant in the $TT$ case.  However,  it is also clear that the size of the corresponding  cross section  is considerably smaller than in the $LL$ case.  From \figref{TT2}  we see that the most relevant coefficient in the $TT$ case is $a_{HHWW}$.  Again, to understand the hierarchy among the coefficients we present next the behaviour of the expansion in powers of $s$ that we have found for the $\amp^{(4)}(W_TW_T \to HH)$:
\begin{figure}[t!]
\begin{center}
\hspace{-3mm}\includegraphics[scale=0.46]{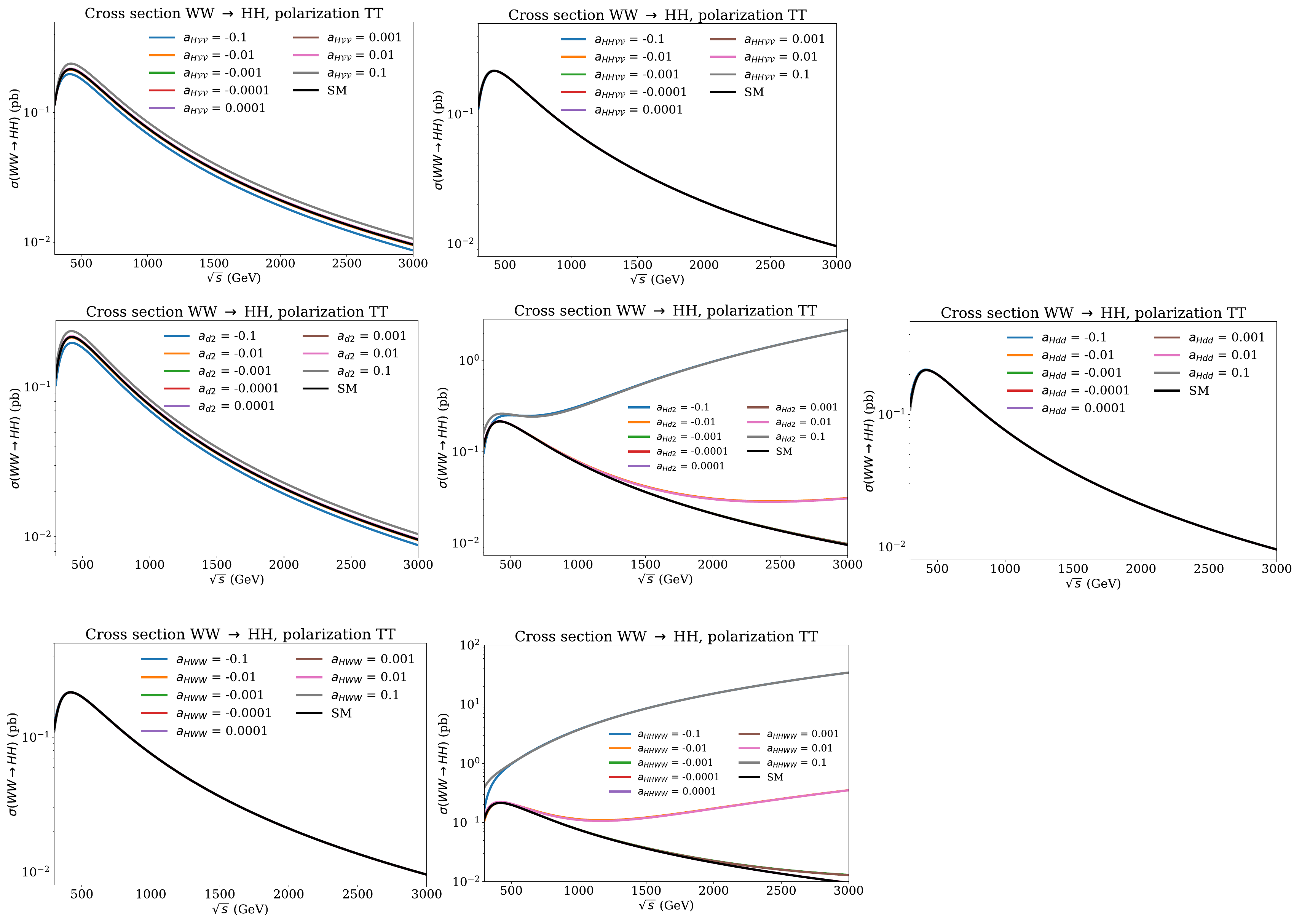} 
\caption{Cross section prediction in the HEFT at subprocess level corresponding to the polarization state $TT$  for different parameter values of $a_{H\mV\mV}$, $a_{HH\mV\mV}$, $a_{d2}$, $a_{Hd2}$, $a_{Hdd}$, $a_{HWW}$ and $a_{HHWW}$,  displayed from left to right and from upper to lower panels,  respectively. The SM prediction (black) is shown for comparison and corresponds to vanishing EChL coefficients.}
\label{TT2}
\end{center}
\end{figure}
\begin{figure}[h!]
\begin{center}
\hspace{-3mm}\includegraphics[scale=0.46]{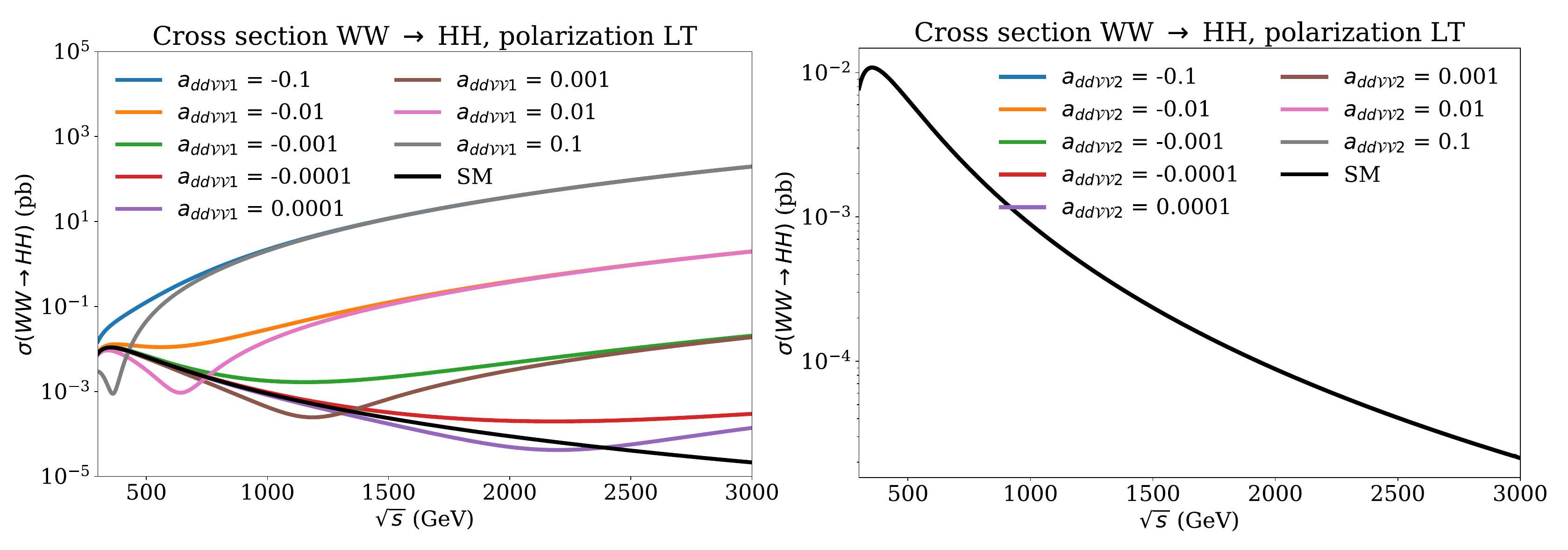} 
\caption{Cross section prediction in the HEFT at subprocess level corresponding to the polarization state $LT+TL$ for different parameter values of $a_{dd\mV\mV1}$ (left) and $a_{dd\mV\mV2}$ (right). The SM prediction (black) is shown for comparison and corresponds to vanishing EChL coefficients.}
\label{LT1}
\end{center}
\end{figure}
\begin{itemize}
\item[1)] The contributions from $a_{dd\mV \mV 1}$, $a_{dd\mV\mV 2}$, $a_{Hd2}$ and $a_{HHWW}$ grow as $\sim s^1$
\item[2)] The contributions from  $a_{d2}$, $a_{Hdd}$, $a_{HWW}$, $a_{H \mV \mV}$ and $a_{HH\mV \mV}$ go as $\sim s^0$
\end{itemize} 
And this explains the behaviour of the cross sections in \figrefs{TT1}{TT2}.

The results for $\sigma_{LT+TL}$  are shown in \figrefs{LT1}{LT2} where  the order chosen in the plots are as before.
In this case, we observe that there are several cases where the coefficients do not affect at all to the cross section and the result is indistinguishable from the corresponding SM prediction.  The most relevant coefficient in this case is  $a_{dd \mV \mV 1}$.  The hierarchy found in the behaviour of the expansion with energy of  $\amp^{(4)}(W_LW_T \to HH)$ is:
\begin{itemize}
\item[1)] The contributions from $a_{dd\mV \mV 1}$ grow as $\sim s^{3/2}$
\item[2)] The contributions from $a_{d2}$, $a_{H \mV \mV}$ and $a_{HWW}$ grow as $\sim s^{1/2}$
\item[3)] The contributions from $a_{dd\mV\mV 2}$,  $a_{Hd2}$, $a_{Hdd}$, $a_{HHWW}$ and $a_{HH\mV \mV}$ vanish.  It is because $\epsilon_+. \epsilon_-=0$ for these $LT$ and $TL$ cases,  thus,  these coefficients do not contribute to $\sigma_{LT+TL}$. 
\end{itemize} 
And this clearly explains the behaviour of the cross sections in \figrefs{LT1}{LT2}.

\begin{figure}[h!]
\begin{center}
\includegraphics[scale=0.46]{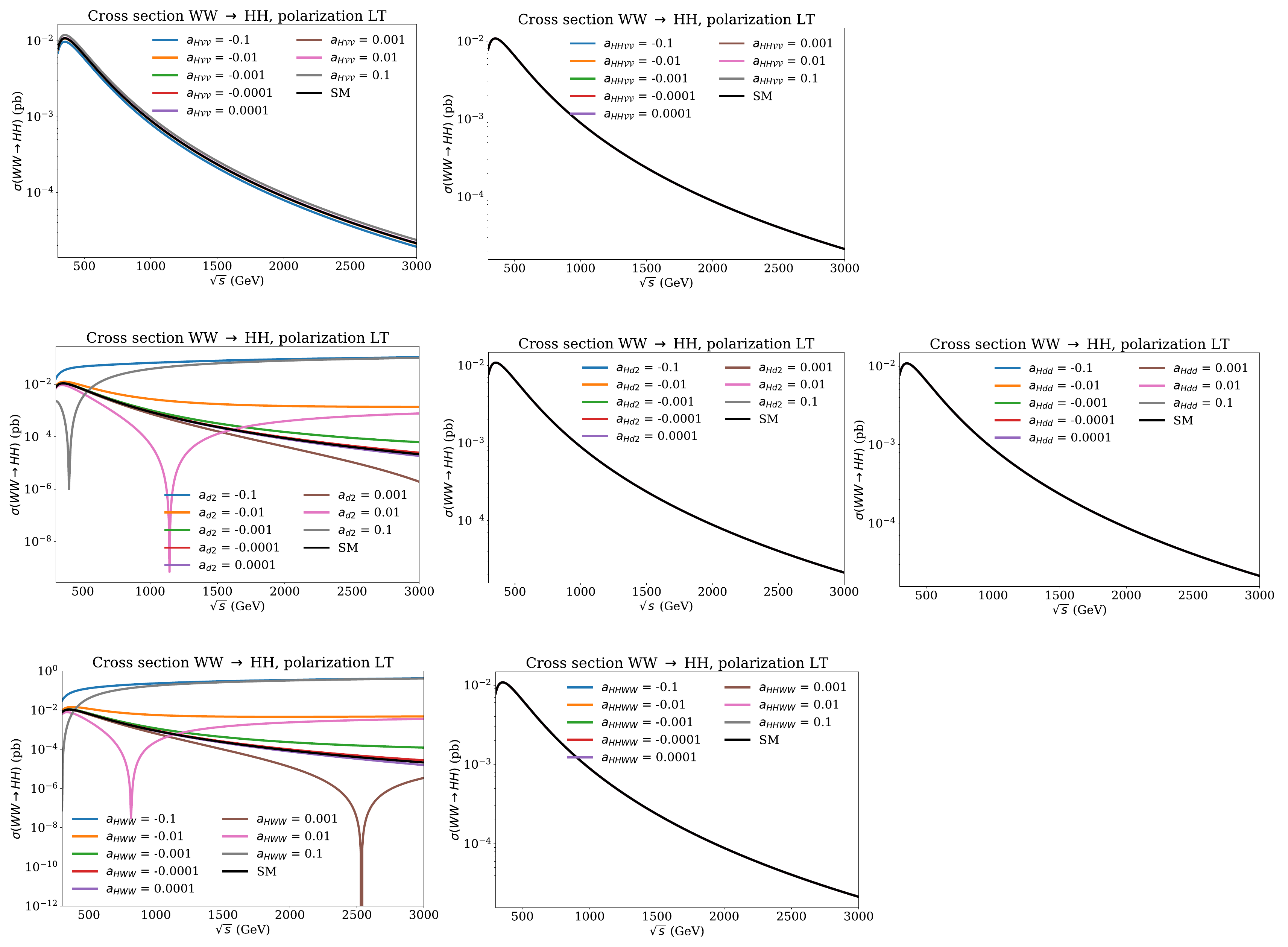} 
\caption{Cross section prediction in the HEFT at subprocess level corresponding to the polarization state $LT+TL$ for different parameter values of $a_{H\mV\mV}$, $a_{HH\mV\mV}$, $a_{d2}$, $a_{Hd2}$, $a_{Hdd}$, $a_{HWW}$ and $a_{HHWW}$,  displayed from left to right and from upper to lower panels,  respectively. The SM prediction (black) is shown for comparison and corresponds to vanishing EChL coefficients.}
\label{LT2}
\end{center}
\end{figure}

\begin{figure}[h!]
\begin{center}
\includegraphics[scale=0.5]{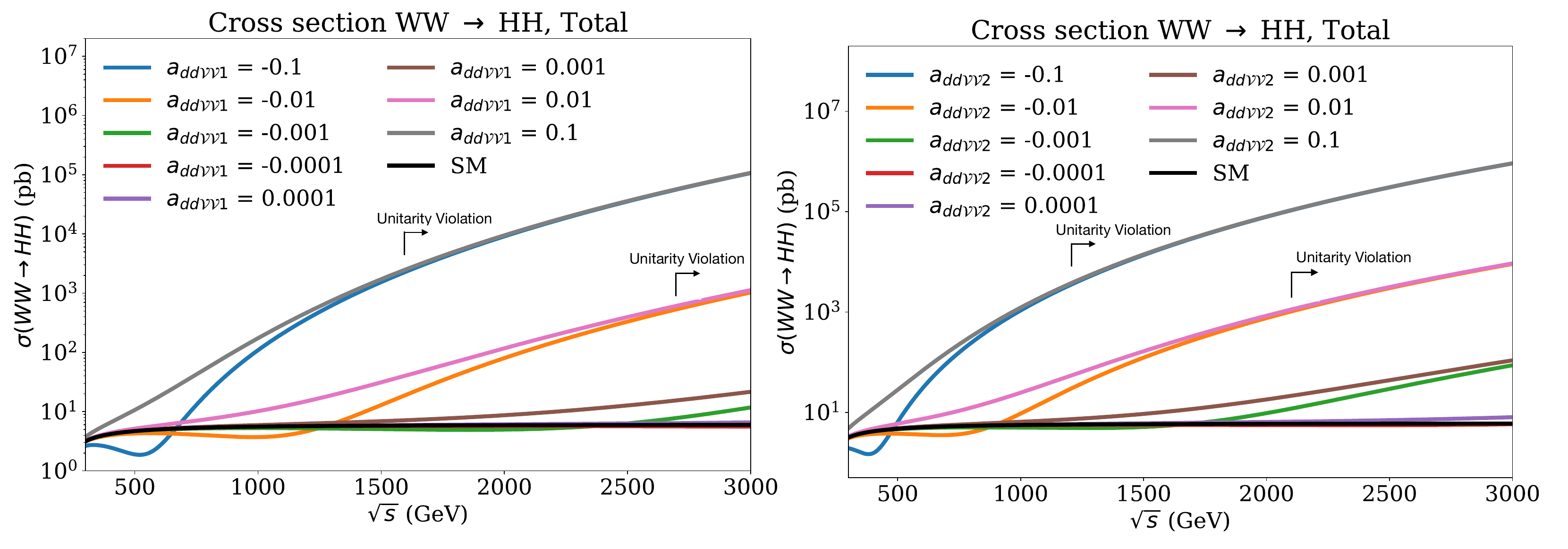} 
\caption{Total unpolarized cross section prediction in the HEFT at subprocess level for different parameter values of $a_{dd\mV\mV1}$ (left) and $a_{dd\mV\mV2}$ (right). The SM prediction (black) is shown for comparison and corresponds to vanishing EChL coefficients. The energy for which unitarity is broken is also shown.  Notice that it only occurs in these plots for parameter values of $\pm0.1$ and $\pm0.01$}
\label{Total-Unpolarized}
\end{center}
\end{figure}

Finally, we show in \figref{Total-Unpolarized} the predictions of the total (unpolarized) cross section as a function of the center-of-mass energy for the two most relevant coefficients $a_{dd \mV \mV 1}$ and $a_{dd \mV \mV 2}$.  Comparing these two plots with the corresponding ones of $\sigma_{LL}$ in \figref{LL1} one can see that the contributions from the $LL$  modes explains fully the pattern with $\sqrt{s}$ and the size of the total cross section (with a factor of (1/9) difference due to the average over the 9 helicity combinations).  Notice also that we have included in this figure the points in energy where the unitarity border is crossed.  For the studied interval in energy here,  this crossing into the unitarity violating region occurs only in the $LL$ channel, only for the largest studied values of  $a_{dd\mV \mV 1}$ and/or  $a_{dd\mV \mV 2}$,  and it is characterized by its dominant $J=0$ partial wave,  $|a_0(s)|$,   crossing above one for that signaled energy. 
 
In summary,  we have shown in this section that the total cross section of this scattering process $WW \to HH$ is dominated by he longitudinal modes and the BSM Higgs physics in the HEFT is mainly determined by the two EChL coefficients $a_{dd \mV \mV 1}$ and $a_{dd \mV \mV 2}$.

\section{$WW\to HH$ in SMEFT}
\label{sec:SMEFT}
\subsection{The relevant SMEFT Lagrangian}\label{sec:SMEFTL}

The SMEFT~\cite{Brivio:2017vri} is built upon the same field content and the same linearly realized $SU(3)\times SU(2)_L\times U(1)_Y$ symmetry as the SM. Contrarily to the HEFT, the Higgs boson is embedded in a $SU(2)_L$ doublet,
%
\begin{equation}
    \phi = \begin{pmatrix} -\text{i} \omega^+ \\
    \frac{H + v + \text{i} \omega^0}{\sqrt{2}}
    \end{pmatrix}\,,
\end{equation}
that is normalized such that the Higgs mass is $m_H^2 = 2\lambda v^2$.

Assuming lepton and baryon number conservation, the SMEFT Lagrangian takes the form:
\begin{equation}
    \mathcal{L}_{\rm SMEFT} = \mathcal{L}_{\rm SM} + \mathcal{L}_6 + \mathcal{L}_8 + \dots\,,~~~\text{with}~\mathcal{L}_d = \frac{a_i}{\Lambda^{d-4}} \mathcal{O}_i^{(d)}
\end{equation}
and $\mathcal{O}_i^{(d)}$ denoting a gauge invariant operator with mass dimension $d>4$. 
The complete non-redundant basis of dim 6 operators was presented in Ref.~\cite{Grzadkowski:2010es}, while that of dim 8 became available only recently~\cite{Murphy:2020rsh,Li:2020gnx}. 
In the Lagrangian above, the suppression by $d-4$ powers of the cut-off scale naturally implies that operators with $d=6$ are LO corrections to the SM Lagrangian, $d=8$ are NLO corrections, and so on. However, depending on the physical problem, there can be cases when the higher-dimensional operators become more relevant while making sense of the SMEFT expansion~\cite{Contino:2016jqw}.

The primary goal of this section is to relate the operators in the SMEFT with the most relevant operators in the HEFT contributing to $WW\to HH$ at large $\sqrt{s}$.  Here again,   large $\sqrt{s}$ means energies at the TeV domain.  Hence, we focus on operators that affect mostly the longitudinal amplitude and lead to the largest growth of the latter with $s$. 
At dim 6, these are:
\begin{equation}
\mathcal{L}_6 =  
\frac{a_{\phi \Box}}{\Lambda^2} (\phi^\dagger \phi) \Box (\phi^\dagger \phi) + \frac{a_{\phi D}}{\Lambda^2} (\phi^\dagger D_\mu \phi) ((D^\mu \phi)^\dagger \phi)\,; 
\label{eq:L6}
\end{equation}
%
%
%
while the relevant dim 8 SMEFT Lagrangian is:
%
%
%
\begin{align}
    \mathcal{L}_8 & = \frac{a_{\phi^6}^{(1)}}{\Lambda^4}(\phi^\dagger \phi)^2 (D_\mu \phi^\dagger D^\mu \phi) +\frac{a_{\phi^6}^{(2)}}{\Lambda^4} (\phi^\dagger \phi) (\phi^\dagger \sigma^I \phi)  (D_\mu \phi^\dagger \sigma^I D^\mu \phi)\\
    & ~~+ \frac{a_{\phi^4}^{(1)}}{\Lambda^4}  (D_\mu \phi^\dagger D_\nu \phi)  (D^\nu \phi^\dagger D^\mu \phi) + \frac{a_{\phi^4}^{(2)}}{\Lambda^4}  (D_\mu \phi^\dagger D_\nu \phi)  (D^\mu \phi^\dagger D^\nu \phi) + \frac{a_{\phi^4}^{(3)}}{\Lambda^4} (D_\mu \phi^\dagger D_\mu \phi)  (D^\nu \phi^\dagger D^\nu \phi)\,. \nonumber
\end{align}
The two-derivative dim 8 operators in the Lagrangian above give the same effects to the scattering process of our interest as the dim 6 operators but their contributions are further suppressed by $\mathcal{O}(\vev^2 /\Lambda^2)$. Therefore, the former are neglected in our analysis.

For purposes of illustration, given the different power counting in the SMEFT and the possibly non-negligible contribution to the total cross section, we carry $\mathcal{O}_{\phi W} \equiv (\phi^\dagger \phi) W_{\mu\nu}^a W^{a\mu\nu}$ in the analysis but do not consider higher dimensional operators containing field strengths.


\subsection{Scattering amplitude in SMEFT}~\label{sec:ampSMEFT}

Analogously to section~\ref{sec:ampHEFT}, below we present the tree level amplitude for the scattering process of our interest from the relevant SMEFT Lagrangian,
\begin{equation}
\mathcal{A}(WW\to HH)_{\rm SMEFT} = \mathcal{A}_{\rm SM}+ \mathcal{A}^{[6]} +\mathcal{A}^{[8]}\,,
\end{equation}
where $\mathcal{A}_{\rm SM}$ is the SM contribution defined in equation~\ref{ampWWtoHH-SM}. The superscripts in squared brackets in the expression above refer to the canonical dim 6 and dim 8 contributions for which the various channels read:
\bear
\amp^{[6]}\vert_S &=& \frac{g^2}{4}\frac{\vev^2}{\Lambda^2}\delta a_{\phi D}\frac{S+8\mh^2}{S-\mh^2}\epsilon _+ \cdot \epsilon _-
+ 6\frac{\vev^2}{\Lambda^2} a_{\phi W}\frac{\mh^2}{\vev^2}\frac{2 \epsilon _-\cdot p_+ \epsilon _+\cdot p_--\left(S-2 \mw^2\right)\epsilon _+\cdot \epsilon _-}{S-\mh^2}\,;\nn\\
\amp^{[6]}\vert_T &=& \frac{g^2}{2}\frac{\vev^2}{\Lambda^2}\delta a_{\phi D}\frac{\mw^2 \epsilon _+\cdot \epsilon _-+\left(\epsilon _-\cdot p_+-\epsilon _-\cdot k_1 \right) \epsilon _+\cdot k_1 }{T-\mw^2}
\nn\\
&& + 2g^2\frac{\vev^2}{\Lambda^2}a_{\phi W}\frac{\epsilon _+\cdot \epsilon _- \left(-\mh^2+\mw^2+T\right)-\epsilon _-\cdot k_1  \epsilon _+\cdot p_-+\epsilon _-\cdot p_+ \left(\epsilon _+\cdot p_-+\epsilon _+\cdot k_1 \right)}{T-\mw^2}\,;\nn \\
\amp^{[6]}\vert_U &=& \amp^{(6)}\vert_{T} \,\,\,{\rm with}\,\,\,T\to U \,\,\,{\rm and}\,\,\, k_1\leftrightarrow k_2  \nn\\
\amp^{[6]}\vert_C &=& \frac{g^2}{4}\frac{\vev^2}{\Lambda^2}\delta a_{\phi D}\epsilon _+ \cdot \epsilon _- +\frac{\vev^2}{{\Lambda ^2}}a_{\phi W}\frac{1}{\vev^2} \left(-2 \left(S-2 m_W^2\right)\epsilon _+\cdot \epsilon _-+4 \epsilon _-\cdot p_+ \epsilon _+\cdot p_-\right)\,;
\nn\\
\amp^{[8]}\vert_C &=& -\frac{g^2}{4}\frac{\vev^2}{\Lambda^4}\left( (a_{\phi^4}^{(1)}+a_{\phi^4}^{(2)}) \left(\epsilon _-\cdot p_+ \epsilon _+\cdot k_1+\epsilon _- \cdot k_1 \left(\epsilon _+ \cdot p_- -2 \epsilon _+ \cdot k_1\right)\right)+a_{\phi^4}^{(3)} \epsilon _+ \cdot \epsilon _- \left(S-2 m_H^2\right)\right) \nn\\
\label{eq:smeftamp}
\eear
up to $\mathcal{O}(a_{\phi W}^2,\delta a_{\phi D}^2/\Lambda^4)$ terms that we have omitted for simplicity. The kinematic variables are defined as in section~\ref{sec:ampHEFT} and momentum conservation 
($p_+ +  p_- = k_1+k_2$) has being used.

The expressions above are obtained after normalizing canonically the fields, as several of the Wilson coefficients contribute to the kinetic terms of the Higgs and the gauge bosons~\cite{Hays:2018zze,Helset:2020yio}. 
The corresponding field redefinitions produce vertices which were zero in the EFT before the rotations. Namely, the coefficients $a_{\phi D}$ and $a_{\phi \Box}$ contribute only to the triple Higgs vertex before canonical normalization while after they are manifest in all the vertices relevant to the process, producing contributions to all $S,T,U$ and $C$ channels. Moreover, such coefficients appear always in the same combination $\delta a_{\phi D} \equiv 4 a_{\phi \Box} - a_{\phi D}$. 
%

Moreover, the effective operators under study give corrections to some of the EW inputs in the set $\{\alpha_{\rm em},\,m_Z^2,\,G_F,\,m_H^2\}$. 
We absorb these corrections by redefining the gauge and the Higgs couplings; 
 therefore, all the parameters in the previous expressions are to be understood as barred parameters, \textit{e.g.} $g\to \overline{g} = (1+\delta g) g$. The explicit rotations are obtained following Ref.~\cite{Hays:2020scx} in order to produce the plots in section~\ref{sec:xsSMEFT}.


Note that the dim 8 four-derivative operators contribute solely to the $WWHH$ vertex hence to the contact amplitude. These and the two-derivative operators are accompanied by different energy dependencies and hence contribute differently to the cross section.

\subsection{Cross section results  in SMEFT}~\label{sec:xsSMEFT}

In this subsection, we present the numerical results for the cross section of $W W \to H H $ sourced by the SMEFT Lagrangian presented in section~\ref{sec:SMEFTL}. We have focused on the operators contributing mostly to the $LL$  modes which are expected to give the largest contributions to the process under study, following the results obtained in previous sections. 

Indeed, performing an expansion of the amplitude $\mathcal{A}^{[6]} +\mathcal{A}^{[8]}$ in powers of $s$, we have found the following behaviour for the highest $\mathcal{O} (s^n)$ terms in $\mathcal{A}(W_L W_L \to HH)$:
\begin{itemize}
\item[1)] The contributions from $a_{\phi^4}^{(1)},\,a_{\phi^4}^{(2)}$ and $a_{\phi^4}^{(3)}$ grow as $\sim s^2$\,;
\item[2)] The contributions from $a_{\phi D}$ and $a_{\phi \Box}$ grow as $\sim s^1$\,;
\item[3)] In comparison, the contribution from $a_{\phi W}$ goes as $\sim s^0$\,.
\end{itemize}
Considering the same expansion in $\mathcal{A}(W_T W_T \to HH)$, we have found:
\begin{itemize}
\item[1)] The contributions from $a_{\phi^4}^{(1)},\,a_{\phi^4}^{(2)}$ and $a_{\phi^4}^{(3)}$ grow as $\sim s^1$\,;
\item[2)] The contributions from $a_{\phi D}$ and $a_{\phi \Box}$ grow as $\sim s^0$\,;
\item[3)] In comparison, the contribution from $a_{\phi W}$ goes as $\sim s^1$\,.
\end{itemize}
Concerning $\mathcal{A}(W_L W_T \to HH)$, the hierarchy found in the behaviour of the expansion with energy is:
\begin{itemize}
\item[1)] The contributions from $a_{\phi^4}^{(1)}$ and $a_{\phi^4}^{(2)}$ grow as $\sim s^{3/2}$\,;
\item[2)] The contributions from $a_{\phi D}$ and $a_{\phi \Box}$ decay as $s^{-1/2}$\,;
\item[3)] The contribution from $a_{\phi^4}^{(3)}$ vanishes\,;

\item[4)] In comparison, the contribution from $a_{\phi W}$ grows as $s^{1/2}$\,.
\end{itemize}
\begin{figure}[b!]
    \centering
    \includegraphics[scale=0.82]{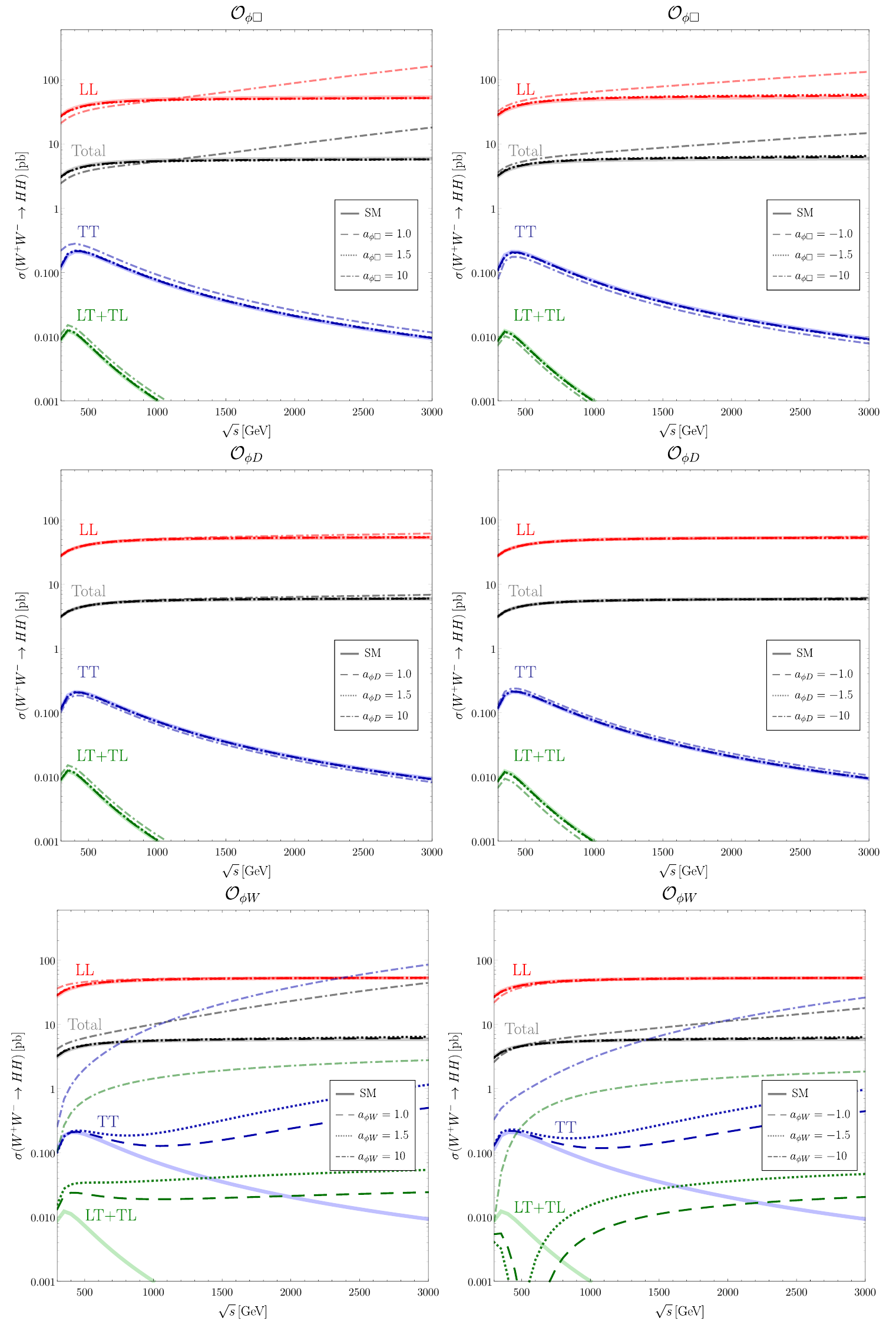}    
    \caption{Contributions to the process $WW\to HH$ in the SMEFT from dim 6  operators for $\Lambda=3.5$ TeV and different values of the Wilson coefficients. The black, red, blue and green colors show the contributions to $\sigma_{\rm tot},\,\sigma_{LL},\,\sigma_{TT}$ and $\sigma_{LT+TL}$, respectively.  The SM prediction is also included for comparison.}
    \label{fig:d6smeft}
\end{figure}

These results explain the behaviour of the cross sections in \figrefs{fig:d6smeft}{fig:d8smeft}, where the BSM results for $\sigma_{LL},\,\sigma_{TT}$ and $\sigma_{LT+TL}$ as a function of $\sqrt{s}$ are shown. In each plot, we explore the effect of each coefficient $a_i$ separately, setting the others to zero values.  In all these plots we assume the cut-off scale $\Lambda = 3.5$ TeV and $a_i=\{\pm 1.0,\,\pm1.5,\,\pm 10,\,\pm 100\}$.
In all plots of these figures the SM predictions are also included.

The large input values chosen above point already to the fact that large BSM Wilson coefficients in the SMEFT are required to see non-negligible deviations from the SM prediction. 
%
Focusing at first on the dim 6 derivative interactions, it can be seen in \figref{fig:d6smeft} that the respective contributions can dominate over the SM in a range of energy $1 \,\text{TeV}\lesssim \sqrt{s} < \Lambda$, but only for $a_{\phi \Box},a_{\phi D} \gtrsim \mathcal{O}(1)$. 

For Wilson coefficients close to the unity, the contributions from dim 6 and dim 8 operators are comparable but correcting the SM prediction only slightly. For example, the contributions from $a_{\phi \Box}=-0.5$ and $a_{\phi^4}^{(3)} = 0.5$ to $\sigma_{LL}$, at $\sqrt{s}=3$\,TeV, read $54.4$ and $53.3$\,pb, respectively. In comparison, the SM prediction is $\sigma_{LL}\approx 53.0$ pb.

\begin{figure}[H]
    \centering
    \includegraphics[scale=0.9]{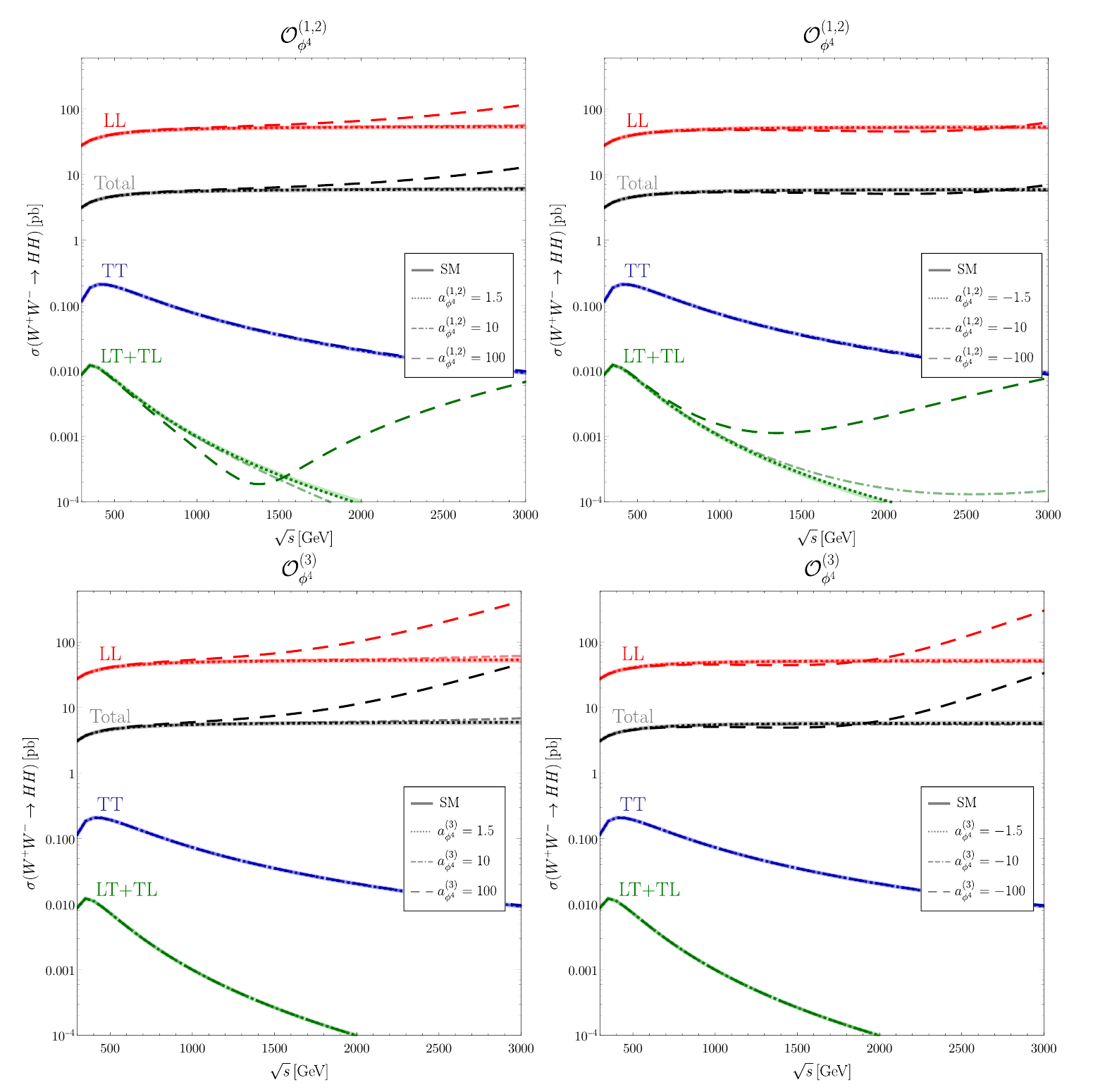} 
    \caption{Contributions to the process $WW\to HH$ in the SMEFT from dim 8 operators for $\Lambda=3.5$ TeV and different values of the Wilson coefficients. The black, red, blue and green colors show the contributions to $\sigma_{\rm tot},\,\sigma_{LL},\,\sigma_{TT}$ and $\sigma_{LT+TL}$, respectively.  The SM prediction is also included for comparison.  As shown in~\cite{Remmen:2019cyz}, the class of operators represented in these figures is subject to tree level positivity constraints. However, these can be spoiled by quantum corrections~\cite{Chala:2021pll}. Hence, in this work we do not take positivity into account and assume that the dim 8 coefficients can have either sign.}
    \label{fig:d8smeft}
\end{figure}

Enlarging $e.g.\,a_{\phi \Box}$, the quadratic terms on the dim 6 coefficients contributing to the cross section eventually start dominating over linear terms and the prediction starts to deviate significantly from the SM. 
Assuming the same numerical values for $a_{\phi^4}^{(i)}$,  and for this large cut-off value of $\Lambda=3.5$ TeV,  the dim 8 contributions are subleading as observed in \figref{fig:d8smeft}.
Note however that depending on the choice for $\Lambda$, the relative size of dim 6 versus dim 8 contributions may change.  All numerical arguments in this discussion follow the results presented in the plots.

As a remark, we point out that the contribution from non-derivative operators like $\mathcal{O}_{\phi W}$ can actually be comparable to that of the derivative operators due to the large enhancement of the $TT$ modes. 
However, in weakly interacting UV theories, $a_{\phi \Box} \gg a_{\phi W}$ \cite{deBlas:2017xtg}.
More importantly, there are strong experimental bounds on the dim 6 coefficients from individual operator at a time or global marginalized fit analyses~\cite{Dawson:2020oco}. Under the former assumption, bounds on $a_{\phi \Box}/\Lambda^2$ require that it is $\mathcal{O}(0.1)\,\text{TeV}^{-2}$ at most, while $a_{\phi D}/\Lambda^2$ and $a_{\phi W}/\Lambda^2$ are bounded to be $\mathcal{O} (0.01)\,\text{TeV}^{-2}$ or smaller, depending on the sign. Such values correspond approximately to $a_{\phi \Box} \sim \mathcal{O}(1) $ and  $a_{\phi D, W} \sim \mathcal{O}(0.1) $ for the cut-off scale of $\Lambda=3.5$ TeV,  leading to only small corrections to the SM prediction.
In the marginalized fit analyses, bounds on these coefficients become weaker and values  $\sim 10$ become allowed. 

On the other hand, bounds on the dim 8 Wilson coefficients allow $a_{\phi^4}^{(i)}/\Lambda^4 \sim 1\,\text{TeV}^{-4}$ (or even larger, depending on the unitarization procedure adopted~\cite{Garcia-Garcia:2019oig}) which is compatible with the largest input shown in \figref{fig:d8smeft}. 
For $a_{\phi^4}^{(i)} > 10$, we observe that the four derivative operators can lead to sizable departures from the SM prediction, specially in bins closer to the cut-off.~

A few other comments are in order concerning the last point. 
First, we verified that when the dim 8 contribution to the cross section dominates over that of the SM, the quadratic terms on the dim 8 coefficient start to take over linear terms in the most energetic bins.
This can occur while making sense of the SMEFT expansion; see for example the discussion in Ref.~\cite{Contino:2016jqw}. 
%
%
Second, even though the dim 8 Wilson coefficients are allowed by data to be larger than the dim 6 Wilson coefficients, we may want to understand if such hierarchy can be accomplished by realistic UV models. This holds trivially if the dim 6 interactions are not generated at tree level by the UV but the dim 8 interactions are. For example, for certain UV theories comprising heavy scalar particles, as that presented in appendix C of Ref.~\cite{Chala:2021pll}, the coefficients of the two dim 6 derivative operators can vanish without making the dim 8 coefficients vanish as well. Moreover, it may happen that both dim 6 derivative operators are generated and non-zero but interfere destructively in the amplitude, such that their contribution vanishes altogether; see section~\ref{sec:ampSMEFT}. 

To summarize, we have identified in this section regimes (experimentally and theoretically consistent) where the total cross cross section of the $WW\to HH$ process is dominated by the longitudinal modes and the BSM Higgs physics in the SMEFT at the TeV energy domain is mainly dominated by the coefficients $a_{\phi^4}^{(1)},\,a_{\phi^4}^{(2)}$ and $a_{\phi^4}^{(3)}$. 
These regimes generically occur for large Wilson coefficients which reflects the proximity to a strongly coupled theory.  For a discussion on the size of the SMEFT coefficients and dimensional/loop  counting rules see,  for instance,  Refs. \cite{Gavela:2016bzc} and \cite{Buchalla:2022vjp}.


%

\section{Comparison HEFT vs SMEFT}
\label{sec:matching}
In this section we compare the previous results for the two EFTs,  HEFT and SMEFT.  We perform the matching of the two theories not at the Lagrangian level, but at the amplitude scattering level.  That means,  in practice,  the following identification:
\begin{equation}
\mathcal{A}(WW\to HH)_{\rm HEFT} =  \mathcal{A}(WW\to HH)_{\rm SMEFT}.
\label{matching-amp}
\end{equation}  
If we split the amplitude in both sides separating explicitly the SM contribution (which cancels in this equation), then this matching can be simply  written as:
\begin{equation}
\mathcal{A}^{(2)}+ \mathcal{A}^{(4)} = \mathcal{A}^{[6]}+ \mathcal{A}^{[8]}. 
\end{equation}
Notice that in this previous equation,  we are  keeping just the linear terms in all the coefficients. 
Notice also that $ \mathcal{A}^{(2)}$ should be re-written after separating the SM part,  and this can be done by considering $\Delta a \equiv 1- a$,  $\Delta b \equiv 1- b$ and $\Delta\kappa_3 \equiv 1- \kappa_3$.  These two features,  then imply replacing,  in $\mathcal{A}^{(2)}$  of \eqref{ampWWtoHH-HEFT}  $(a\kappa_3)$ by $-(\Delta a + \Delta \kappa_3)$,  $a^2$ by $-2 \Delta a,$ and $b$ by $-\Delta b$, and in $\mathcal{A}^{(4)}$  of \eqref{ampWWtoHH-HEFT} setting $a=\kappa_3=1$. 

Finally, the equation of the matching of the amplitudes is solved in terms of the EFT coefficients. This solving takes into account all kinematical structures,  including the dependence in $s$ and $\cos\theta$ with $\theta$ the scattering angle,  the  polarization vector products,  and the mass dependencies.
 We then arrive at the following matching equations among the EFT coefficients:
\bear
 a-1 &=&   \frac{1}{4} \frac{\vev^2 }{\Lambda^2} \delta a_{\phi D}   \nn\\
b -1&=&   \frac{\vev^2 }{\Lambda^2} \delta a_{\phi D}  \nn\\
\kappa_3 - 1&= & \frac{5}{4} \frac{\vev^2 }{\Lambda^2} \delta a_{\phi D}  \nn\\
a_{HWW} &=& -\frac{\vev^2}{2 \mw^2} \frac{\vev^2 }{\Lambda^2} a_{\phi W}  \nn\\
a_{HHWW} &=& - \frac{\vev^2}{4\mw^2}\frac{\vev^2 }{\Lambda^2} a_{\phi W}  \nn\\
a_{dd\mV\mV1} &=& \frac{\vev^4}{4 \Lambda^4} \left[ a_{\phi^4}^{(1)} + a_{\phi^4}^{(2)}\right]  \nn\\
a_{dd\mV\mV2} & = & \frac{\vev^4}{4\Lambda^4}  a_{\phi^4}^{(3)}
\label{eq:matching}
\eear
while $a_{H\mV\mV},\,a_{HH\mV\mV},\,a_{d2},\,a_{d2}$ and $a_{Hdd}$ have no counterpart in the SMEFT (given the reduced set of operators under study).
The results in the first two equations involving $a_{\phi \Box}$ agree with those obtained in Ref.~\cite{Gomez-Ambrosio:2022giw} where the matching was performed at the Lagrangian level.

Some comments on the above relations are in order.  Firstly,  we see that the matching  among the coefficients occurs across different orders of the two expansions,  in chiral and canonical dimensions respectively.  
While the HEFT coefficients $a$,  $b$ and $k_3$,  of chiral dim 2,  are related with the coefficient  $\delta a_{\phi D}$ of canonical dim 6,  the HEFT coefficients  $a_{HWW}$ and $a_{HHWW}$,  of chiral dim 4,  are related with $a_{\phi W}$,  also of canonical dim 6.
 On the other hand,  the HEFT coefficients $a_{dd\mV\mV1}$ and $a_{dd\mV\mV2}$ from chiral dimension 4 are related with $a_{\phi^4}^{(1,2,3)}$ from canonical dimension 8.  Secondly,  in these HEFT/SMEFT relations we detect some correlations.  For instance,   whereas in the HEFT $a_{HWW}$ and $a_{HHWW}$ are independent parameters, they are correlated in the SMEFT by $a_{HWW}=2 a_{HHWW}$.  Similarly, $a$ and $b$ are independent parameters in the HEFT, whereas they are correlated in the SMEFT by $(b-1)=4(a-1)$.  These and other correlations reflect the fact that, in some sense, the SMEFT is contained in the HEFT. 
\begin{table}[t!]
\centering
\begin{tabular}{|c|c|c|c|c|c|}
\cline{2-4}
\multicolumn{1}{c|}{\bf{Matching : HEFT  (SMEFT)}} & $\Lambda=$1 TeV&$\Lambda=$2 TeV& $\Lambda=$3 TeV\\
\hline
\multirow{3}[4]*{$a_{dd\mV\mV1} \left(a_{\phi^4}^{(1)} + a_{\phi^4}^{(2)}\right)$}
&$\pm 0.01$  ($\pm 11$)&$\pm 0.01$ ($\pm 175$) &$\pm 0.01$ ($\pm 885$)
\bigstrut\\\cline{2-4}
&$\pm 0.001$  ($\pm 1.1$)&$\pm 0.001$  ($\pm 17.5 $) & $\pm 0.001$  ($\pm 88.5 $)
\bigstrut\\\cline{2-4}
\hspace{-10mm}$a_{dd\mV\mV2} \left(a_{\phi^4}^{(3)}\right)$&$\pm 0.0001$  ($\pm 0.11$)&$\pm 0.0001$  ($\pm 1.75$) & $\pm 0.0001$  ($\pm 8.85$)
\bigstrut\\
\hline
\end{tabular}
\caption{ Numerical matching among the HEFT and SMEFT coefficients for several choices of the $\Lambda$ cut-off.  We display here just the most relevant HEFT coefficients for the forthcoming study at TeV $e^+e^-$ colliders which are $a_{dd\mV\mV1}=\eta$ and  $a_{dd\mV\mV2}=\delta$.}
\label{table:matching}
\end{table}
On the other hand,  we also see that some NLO effects of the SMEFT cannot be matched to the HEFT if one assumes $a=b=\kappa_3=1$.  For example,  we see in \eqref{eq:matching} that 
it is not possible to match the effect of $a_{\phi \Box}$ alone with HEFT coefficients after imposing that $a=b=\kappa_3 = 1$.



Finally,  to learn on the relative size of the coefficients in the two theories,  we present in \tabref{table:matching} the numerical predictions of the matching relations in \eqref{eq:matching} for the most relevant NLO-HEFT parameters,  $a_{dd\mV\mV1}$ and $a_{dd\mV\mV2}$, and for three possible values of the SMEFT cut-off of $\Lambda=1,2,3$ TeV.  In this table, we clearly see that, in order to get large departures with respect to the SM in the  SMEFT cross sections from the canonical dimension 8 coefficients,  $a_{\phi^4}^{(1,2,3)}$,  being  comparable to those in the HEFT from the chiral dimension 4 coefficients,   $a_{dd\mV\mV1,2}$, one needs rather large SMEFT coefficients,  as already said.  For instance, a value of $a_{dd\mV\mV2} \sim 0.001 $ requires a value of $a_{\phi^4}^{(3)} \sim 1.1 $ for $\Lambda=1$ TeV and  $a_{\phi^4}^{(3)} \sim 88 $ for $\Lambda=3$ TeV. This is compatible with what we have learnt in the previous section by studying numerically the departures, as a function of energy,  of the SMEFT cross section  with respect to  the SM one.  Concretely,  by fixing  $\Lambda=3.5$ TeV  we found relevant departures from the dim 8 operators,  at the TeV energy domain,   if the coefficients are taken as large as 
$\mathcal{O}(100)$,  signaling a strongly underlying interacting UV theory.   In the next section, we will evaluate some phenomenological consequences from these dim 8 operators at $e^+e^-$ colliders with energies in the TeV domain.

\section{Sensitivity to the EFT coefficients at TeV $e^+e^-$  colliders via $HH \nu \bar\nu$ production}
\label{sec:sensitivity}

\begin{figure}[ht!]
	\begin{center}
		\begin{tabular}{cc}
			\centering
			\hspace*{-9mm}
			\includegraphics[scale=0.3]{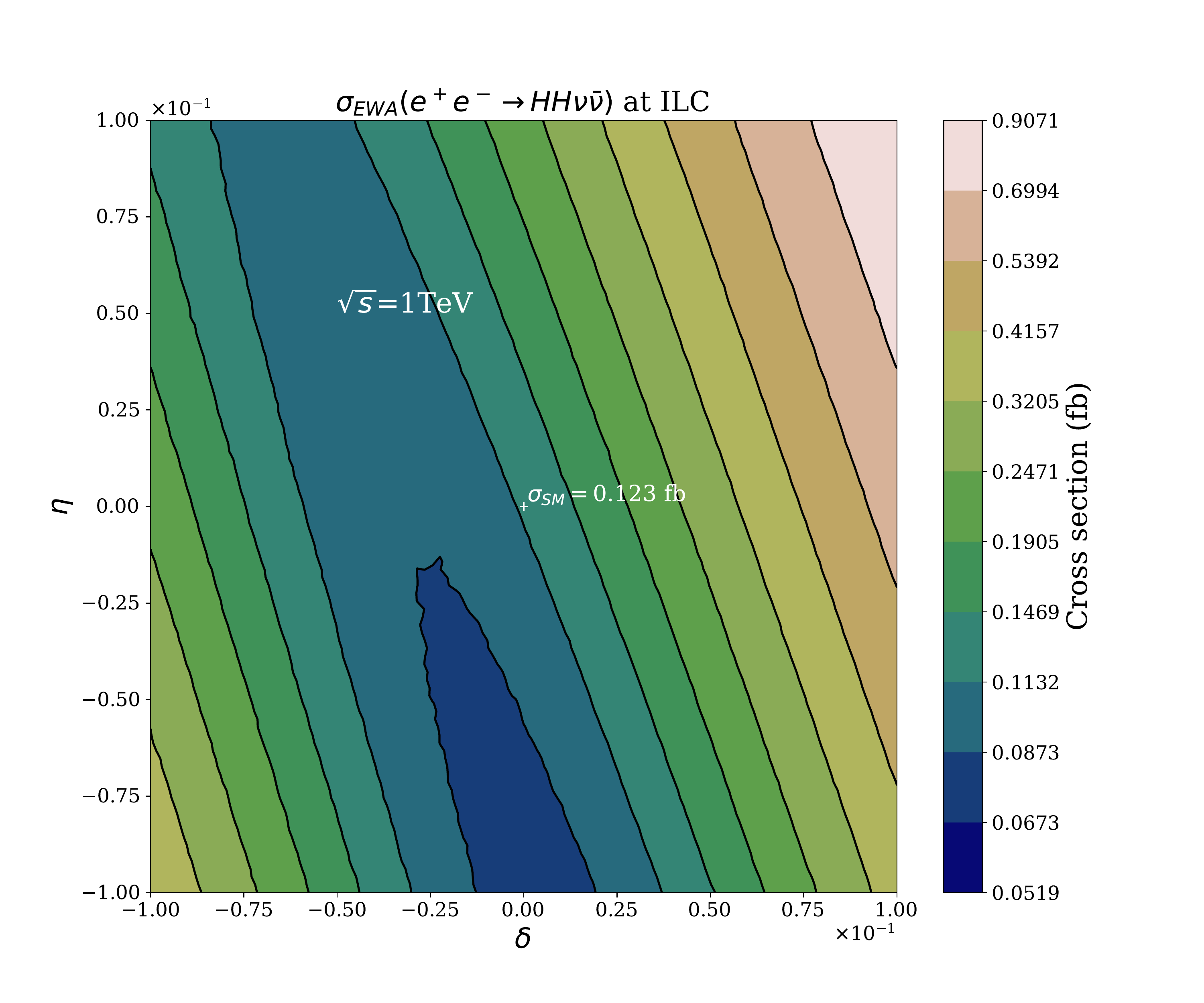} &	\hspace*{-10mm}\includegraphics[scale=0.3]{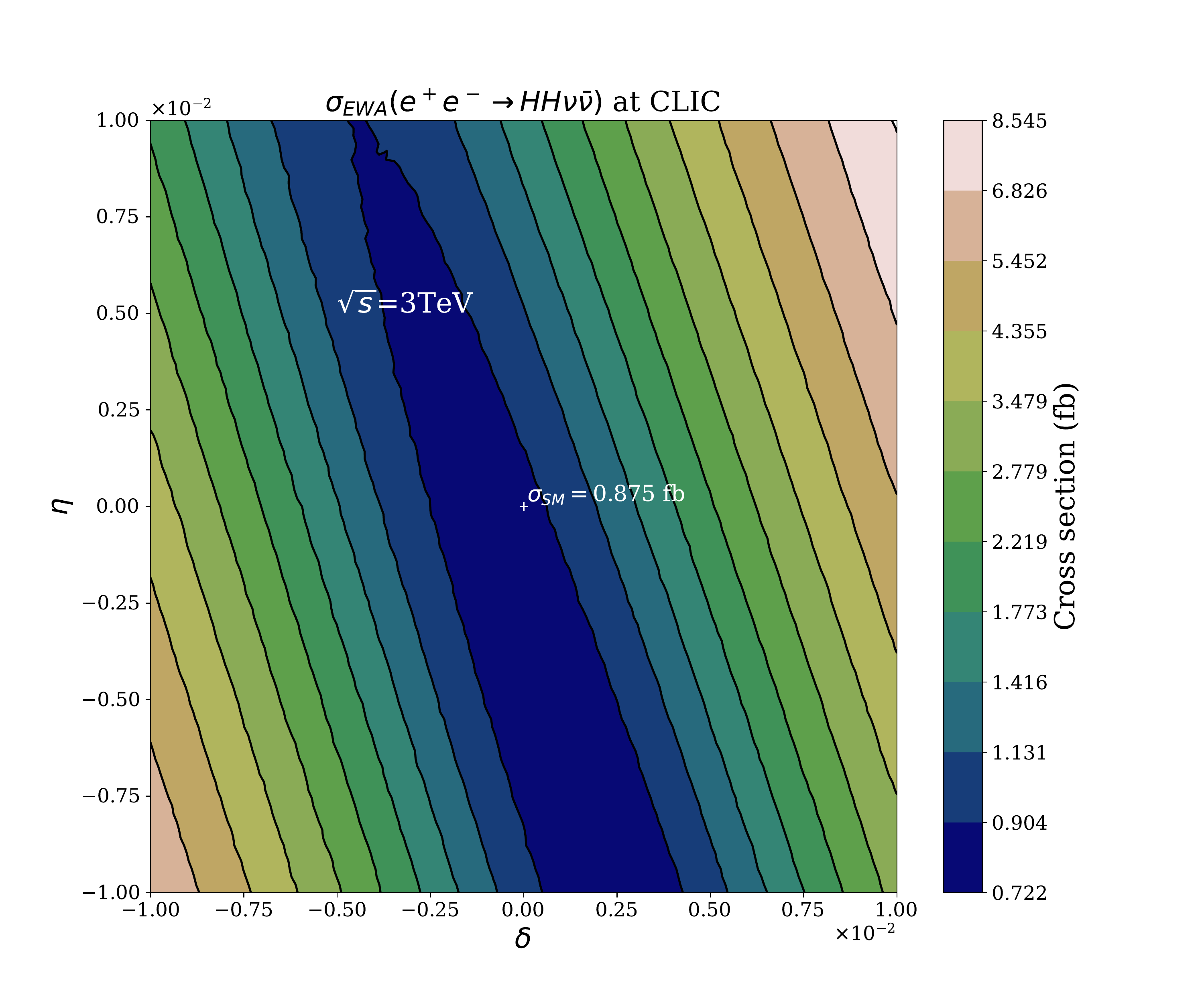}  \\	
			\hspace*{-9mm}\includegraphics[scale=0.3]{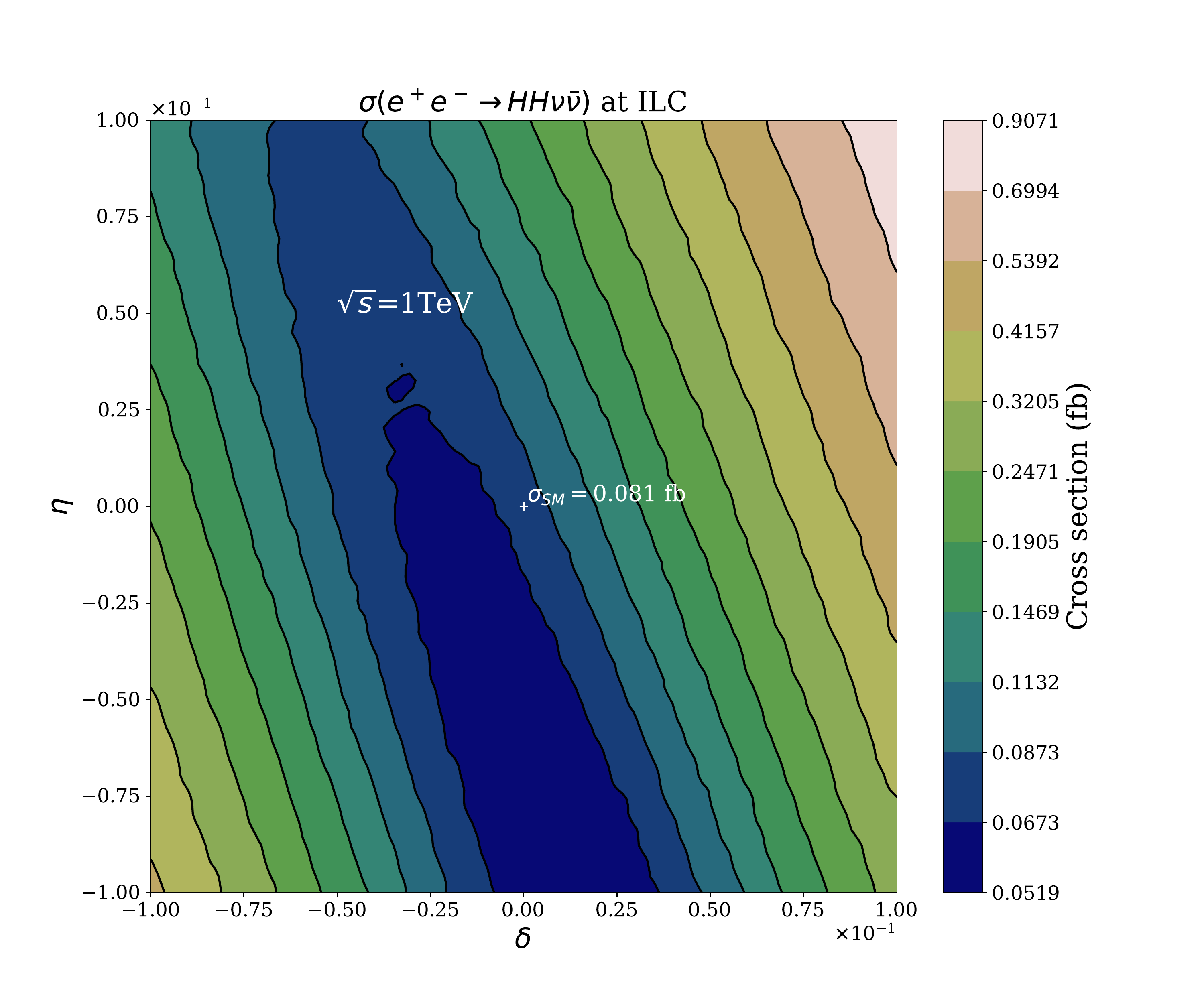} &	\hspace*{-10mm}\includegraphics[scale=0.3]{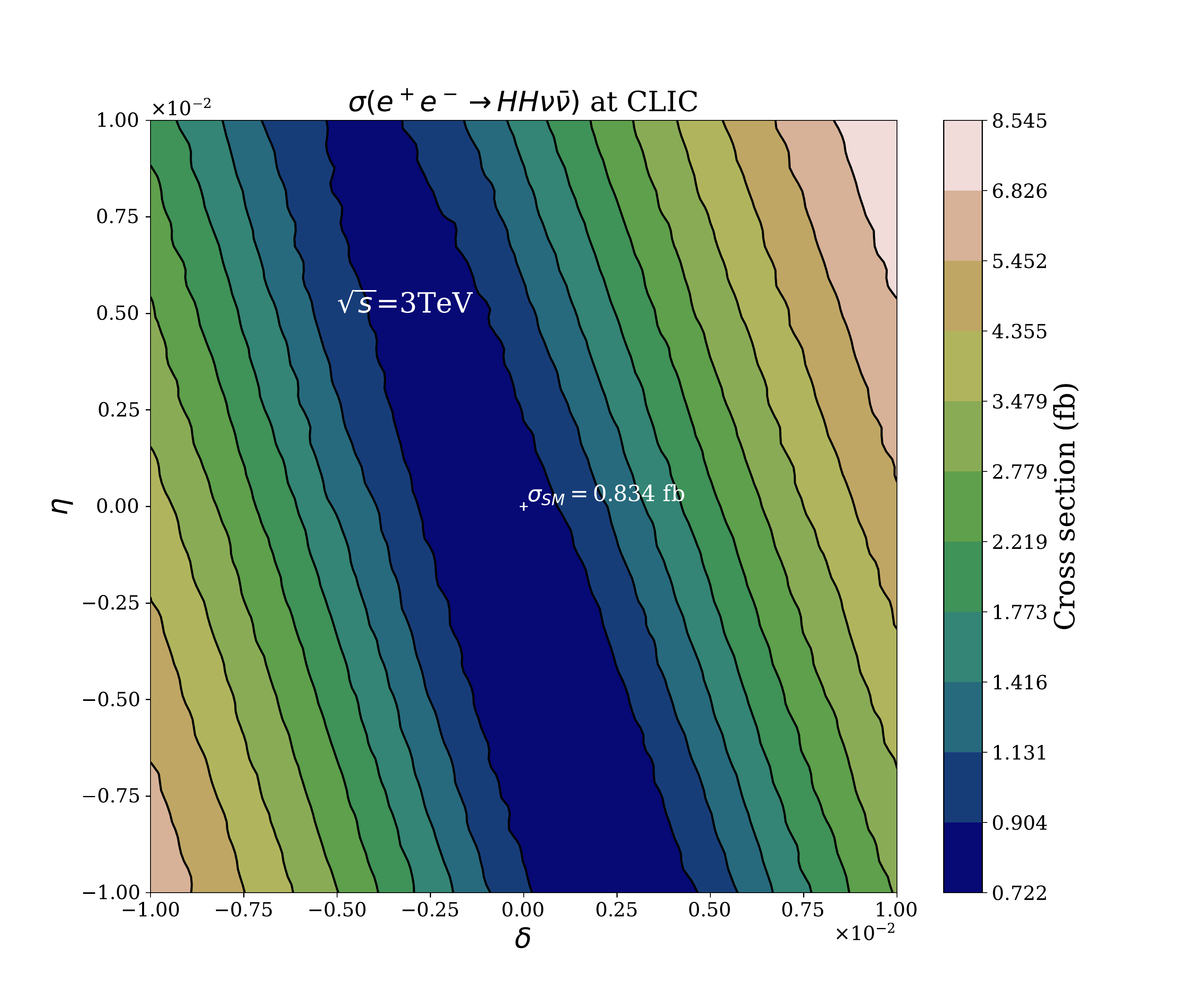}  \\
		\end{tabular}
\caption{EFT predictions for the cross section $\sigma(e^+e^- \to HH \nu \bar \nu)$ in the $(\delta,\eta)$ plane for ILC (left panels) and CLIC (right panels).  The approximate results using the EWA are displayed in upper panels and the full results using MG in the lower panels.  The SM predictions are also included. }
\label{plot-EWAcompHEFT}
	\end{center}
\end{figure}

\begin{figure}[ht!]
\begin{center}
\hspace*{-5mm}\includegraphics[scale=0.3]{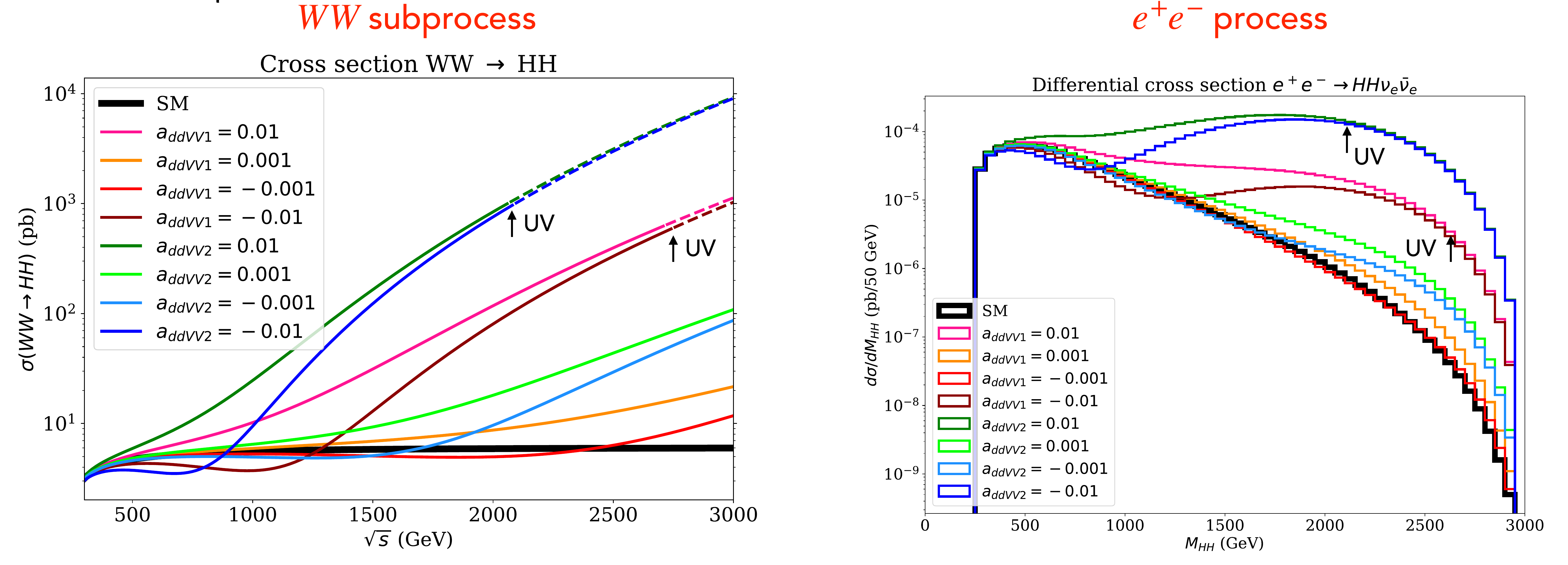}
\caption{Correlation between the enhancements due to BSM physics in the subprocess (left),  $ \sigma(WW \to HH)$,  at large subprocess energy $\sqrt{s}$,  and in the 
$e^+e^-$ process (right),  $d\sigma/dM_{HH}$ at large invariant mass $M_{HH}$,   for various values of the relevant EFT parameters,  $a_{dd\mV\mV1}=\eta$ and $a_{dd\mV\mV2}=\delta$.  Here the CLIC energy 3 TeV is set.  The SM predictions are also included in these plots for comparison.}
\label{dif-xsection}
	\end{center}
\end{figure}

In this section we explore the sensitivity to the EFT BSM Higgs couplings at the future planned TeV $e^+e^-$  colliders.  We first perform a numerical computation of the total cross section at these colliders for the full di-Higgs production process $\sigma(e^+e^- \to HH \nu \bar\nu)$ and later we analyze the sensitivity to the EFT coefficients by considering the particular events with four $b$-jets and missing energy, coming from the dominant $H$ decay into $b\bar b$ pairs, namely,  we analyze the total process 
$e^+ e^- \to b \bar b b \bar b \nu \bar \nu$.  For this analysis we focus on the two most relevant  EFT coefficients,  $\delta$ and $\eta$ of the NLO tree level scattering amplitude, which have been related in the previous sections with the coefficients in both EFTs,  the HEFT and the SMEFT.  The corresponding analysis of the LO tree level HEFT coefficients, $a$, $b$ and $\kappa_3$ was done in Ref.~\cite{Gonzalez-Lopez:2020lpd}, thus we do not repeat it here.  For the present analysis of the NLO coefficients we set the LO ones to their SM values,  i.e.,  in the HEFT case,  we set $a=b=\kappa_3=1$.   

For the computation of the full cross section we use  \textsc{MadGraph5} (MG5) \cite{Alwall:2014hca} which generates and accounts for all the Feynman diagrams contributing to the full scattering process,  $e^+e^- \to HH \nu \bar \nu$.  Therefore,  all the participating diagrams are included, i.e.,  those with $WW$ fusion configuration  and also the others that do not have this configuration,  like  those with intermediate $Z$ bosons which decay to $\nu \bar \nu$,  i.e.,  
$e^+e^- \to  H H  Z \to HH \nu \bar \nu$.  These latter configurations are known to be highly subdominant as compared to the $WW$ fusion ones,  in the case of  $e^+e^-$ colliders with TeV energies.  For a comparison of these two contributions in the SM case,  see for instance
Ref.~\cite{Gonzalez-Lopez:2020lpd}.  In this work,  we focus  on two particular projects with very high energy in the TeV range: 
CLIC \cite{Abramowicz:2016zbo,CLICdp:2018cto, Roloff:2019crr} with $\sqrt{s}=3\, {\rm{TeV}}$  and $L=5\,  {\rm ab}^{-1}$,  and ILC \cite{Strube:2016eje, Bambade:2019fyw} with $\sqrt{s}=1 \,{\rm {TeV}}$ and  $L=8\,  {\rm ab}^{-1}$. 

In order, to check the importance of the $WW$ fusion channel for the present study of the most relevant BSM Higgs coefficients,  
$\delta$ and $\eta$,  and before going to the study of the accessible region to these coefficients at future colliders,  we have compared first the two following cross sections.  On one hand we have determined the full cross section with  MG5,  as already said.  On the other hand we have computed the cross section within the so-called  Effective $W$ Approximation (EWA) where the process is factorized into  the production of two $W$'s that are radiated by the initial electrons and positrons and the subsequent production of the two Higgs bosons by the scattering  subprocess,  $WW \to HH$,  as it is represented generically in \figref{plot-ee}.  The EWA takes into account only the $WW$ fusion contribution to di-Higgs pair production in $e^+e^-$ colliders and, therefore,  by comparing the two results for the cross section from the EWA and from MG5 we will be able to determine quantitatively the relevance of this $WW$ channel. 

In short,  the prediction in the EWA displaying the above mentioned factorisation is given by:
\begin{align}
    \sigma_{\rm EWA}(e^+e^-\to HH \nu \bar \nu) (s) = \int dx_1\int dx_2 \sum_{X,Y} f_{W_X}(x_1) f_{W_Y} (x_2)\; \hat{\sigma}(W_X W_Y\to HH)(\hat{s}),
\end{align}
 where $\sqrt{s}$ is the center of mass energy of the $e^+e^-$ process and $\sqrt{\hat{s}}$ the one of the $WW$ subprocess.   $x_1$ and $x_2$ are the corresponding momentum fractions of the two $W$'s with  respect to the parent fermions.  These $x_{1,2}$ also relate the two center-of-mass energies of the process $\sqrt{s}$ and subprocess $\sqrt{\hat s}$  by $\hat{s} =x_1x_2 s$. The subindices $X,Y$ refer to the polarization of the $W$ bosons (longitudinal or transverse).  Notice that different polarizations must be taken into account separately,  as the probability of radiating a $W$ boson depends on whether it is longitudinally or transversely polarized.  Consequently,  one has to make the convolution of each polarized cross section with the corresponding  distribution functions of the $W$ bosons $f_{W_{L,T}}(x)$.  To compute this cross section $\hat{\sigma}$ we write it  in terms of the polarized amplitudes $ \amp(W_XW_Y \to HH)$ (already presented in the previous sections) which we generate using \textsc{FeynArts-3.10} \cite{FeynArts} and \textsc{FormCalc-9.6} \cite{FormCalc-LT}, and then perform the integration using VEGAS \cite{Lepage:1977sw} and a private PYTHON code.  The analytical expressions that we use for the $W$ distribution functions $f_{W_{L,T}}(x)$ are taken from \cite{Dawson:1984gx} and correspond to the so-called improved EWA, that keeps corrections of order $m_W^2/E^2$, with $E$ being the energy of the parent fermion  radiating the $W$. This improved EWA works better than the most frequently used Leading Log Approximation (LLA) EWA, which is only valid in the very high energy limit, $E\gg m_W$.  The formulas for the LLA-EWA can also be found in  \cite{Dawson:1984gx}. 

The results of this comparison,  $\sigma_{\rm EWA}$ versus $\sigma_{\rm MG5}$,  are shown in \figref{plot-EWAcompHEFT}. These plots display the contourlines for the cross section predictions in the $(\delta, \eta)$ plane. The first (second) row shows  the EWA (MG5) results for the two chosen colliders,  ILC (on the first column) and CLIC (on the second column). 
 The corresponding SM predictions are also included for comparison.  The parameter values explored for these most relevant EFT parameters, $\eta$ and $\delta$,  in these plots  are chosen in the range $[-0.1, +0.1]$ for ILC and $[-0.01, +0.01]$ for CLIC.  
 
 The most important conclusions from these plots are the following.  Firstly,  all the contourlines in the $(\delta, \eta)$ plane display the expected elliptical shape, which can be easily understood from the dependence already shown of the subprocess amplitudes in terms of these two parameters $\eta$ and  $\delta$.  Secondly, the departures of the BSM predictions with respect to the SM ones are quite sizeable, particularly in the upper right corners of these plots,  where a factor of about 10 larger cross sections than in the SM case are obtained.  For instance, for $(\delta, \eta)=(0.1, 0.1)$  we get (with MG5) 
 $\sigma^{\rm BSM}_{\rm ILC}=0.9\, {\rm fb}$ to be compared with  $\sigma^{\rm SM}_{\rm ILC}=0.081\, {\rm fb}$; and for $(\delta, \eta)=(0.01, 0.01)$ we get  $\sigma^{\rm BSM}_{\rm CLIC}=8.54\, {\rm fb}$ to be compared with  $\sigma^{\rm SM}_{\rm CLIC}=0.834\, {\rm fb}$.  Thirdly, the lowest predictions in these plots do not correspond to $(\delta, \eta)=(0,0)$. This means that in some regions of the $(\delta, \eta)$ parameters space  there are 
 negative interferences producing lower predictions  for BSM  than in the SM.  Finally, regarding the $\sigma_{\rm EWA}$ versus $\sigma_{\rm MG5}$ comparison, we find that the EWA is indeed an excellent approximation for CLIC and a quite reasonable approximation for ILC.  The compared rates for the SM case are: $\sigma^{\rm SM}_{\rm EWA}=0.875\, {\rm fb}$ versus $\sigma^{\rm SM}_{\rm MG5}=0.834\, {\rm fb}$ for CLIC; and $\sigma^{\rm SM}_{\rm EWA}=0.123\, {\rm fb}$ versus $\sigma^{\rm SM}_{\rm MG5}=0.081\, {\rm fb}$ for ILC. The convergence of the EWA and the MG5 results are substantially better for the BSM results than for the SM ones, in particular,  in the areas of the $(\delta, \eta)$ plane with the largest cross sections.  For instance,  in the upper right corner of these plots, we find the following results: 1) for $(\delta, \eta)=(0.01,0.01)$ at CLIC we get $\sigma_{\rm EWA}=8.545\,{\rm fb}$ versus $\sigma_{\rm MG5}=8.545\,{\rm fb}$, i.e.,  in full agreement,  2) for $(\delta, \eta)=(0.1,0.1)$ at ILC we get $\sigma_{\rm EWA}=0.907\,{\rm fb}$ versus $\sigma_{\rm MG5}=0.907\,{\rm fb}$, i.e.,  in full agreement again.  The most important conclusion from this good agreement EWA versus MG5 is that the $WW$ fusion channel fully dominates the cross section of $e^+e^- \to HH \nu \bar \nu$. 
 \begin{figure}[ht!]
	\begin{center}
		\begin{tabular}{cc}
			\centering
			\hspace*{-9mm}
			\includegraphics[scale=0.3]{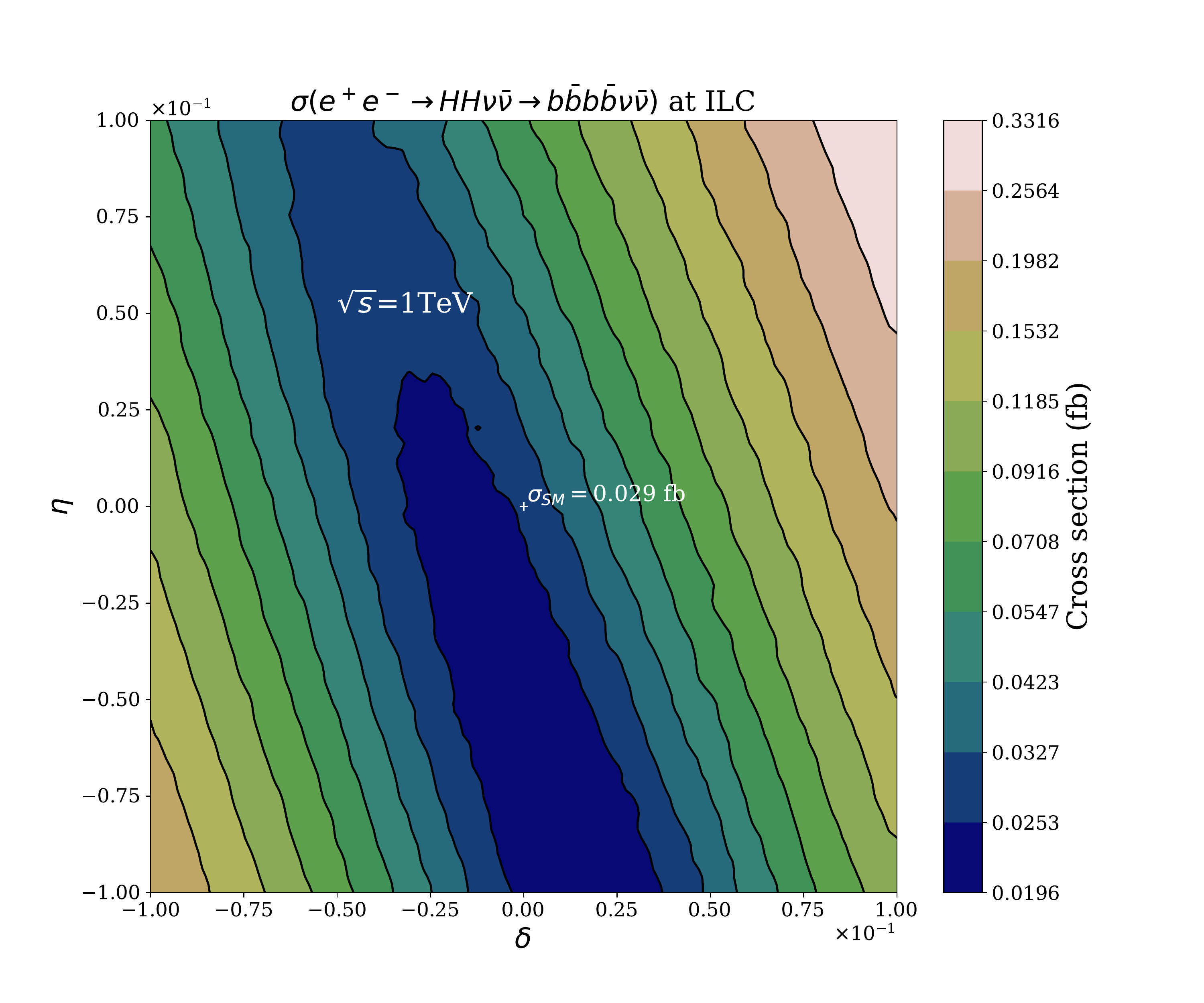} &	\hspace*{-10mm}\includegraphics[scale=0.3]{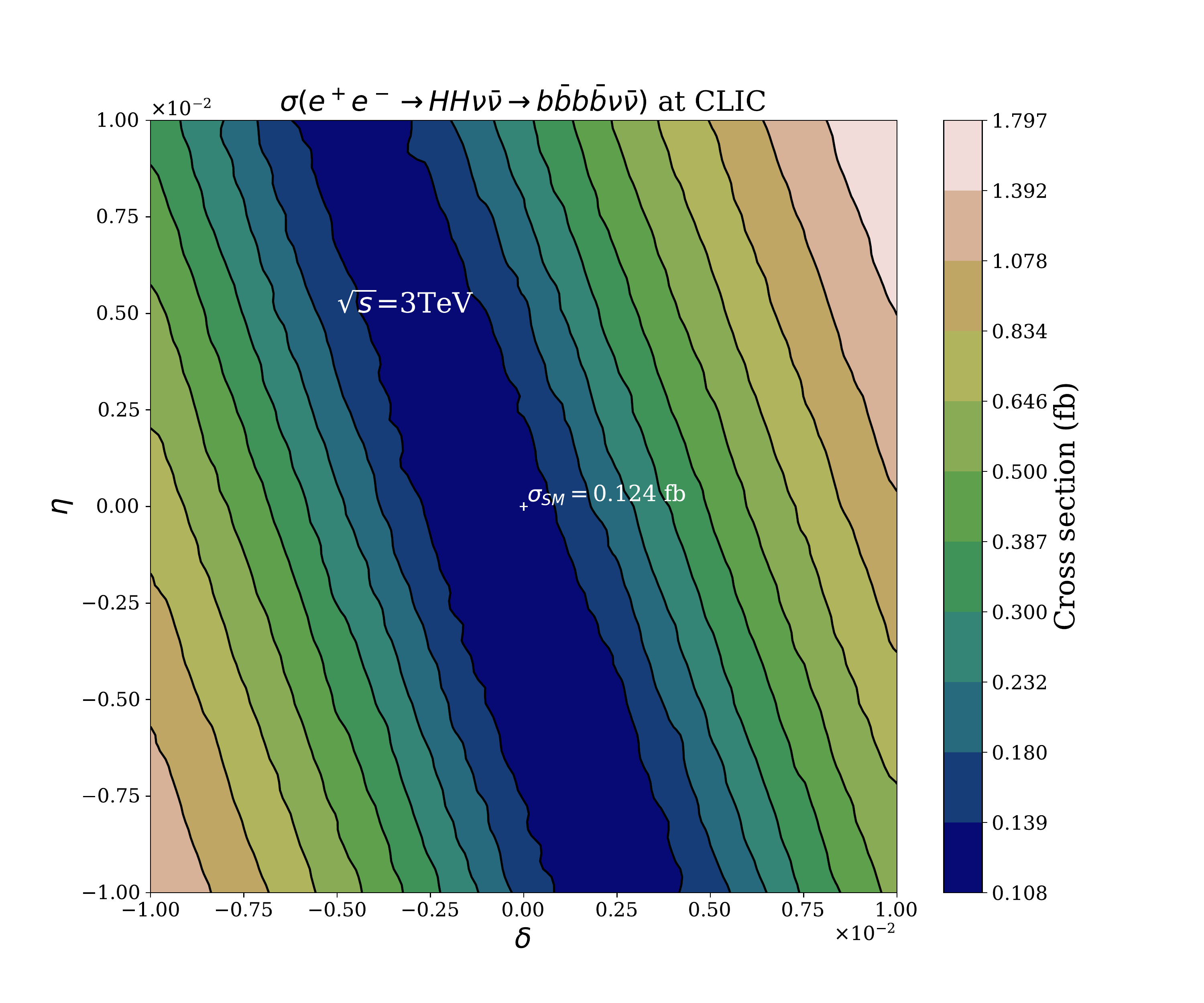}  \\	
		\end{tabular}
\caption{EFT predictions for the cross section $\sigma(e^+e^- \to b \bar b b \bar b  \nu \bar \nu)$ in the 
$(\delta,\eta)$ plane for ILC (left panel) and CLIC (right panel).  Here the cuts and efficiencies described in the text are applied.  The SM predictions are also included.}
\label{plot-xsHEFT}
	\end{center}
\end{figure}

\begin{figure}[ht!]
	\begin{center}
		\begin{tabular}{cc}
			\centering
			\hspace*{-9mm}
			\includegraphics[scale=0.3]{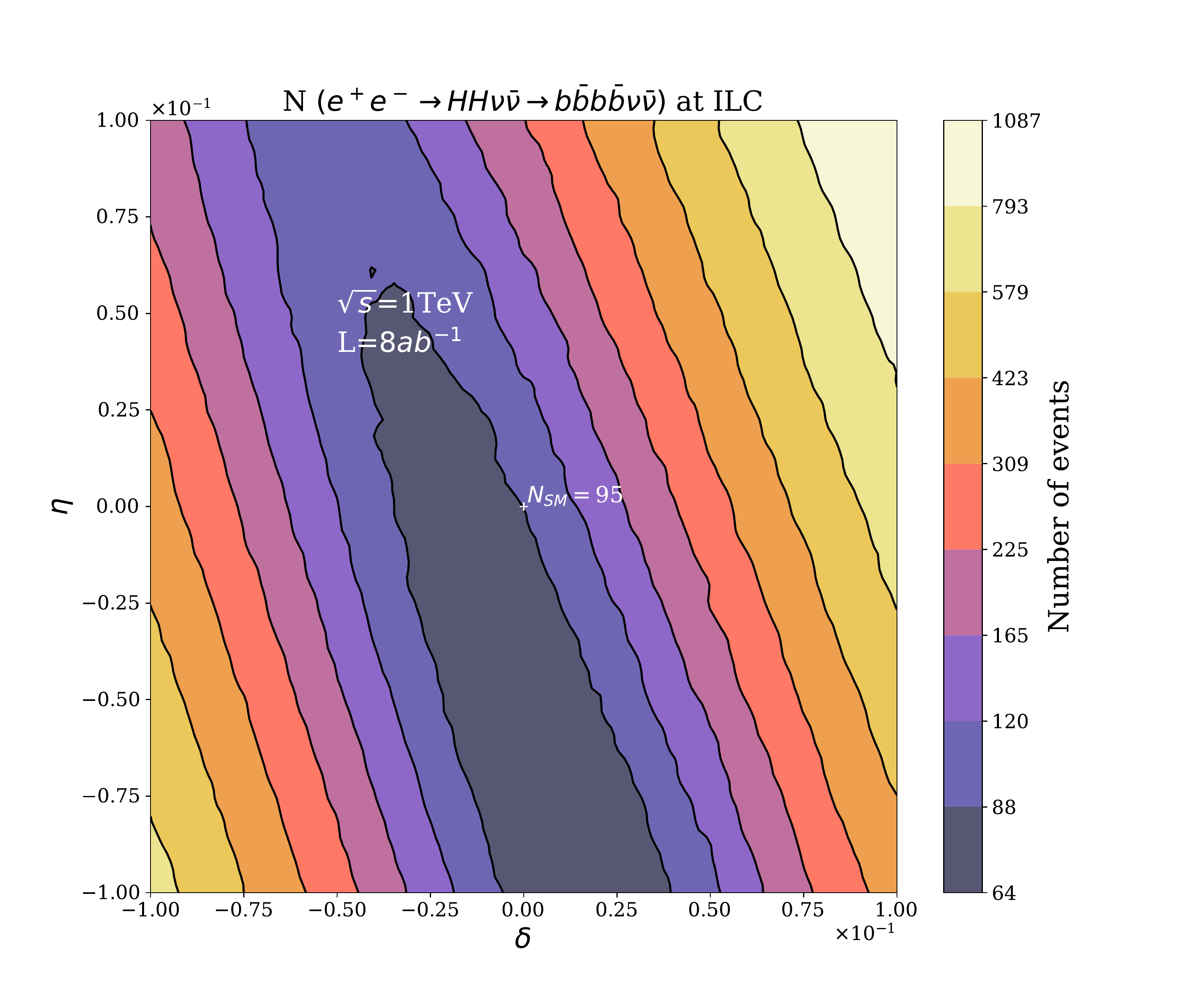} &	\hspace*{-10mm}\includegraphics[scale=0.3]{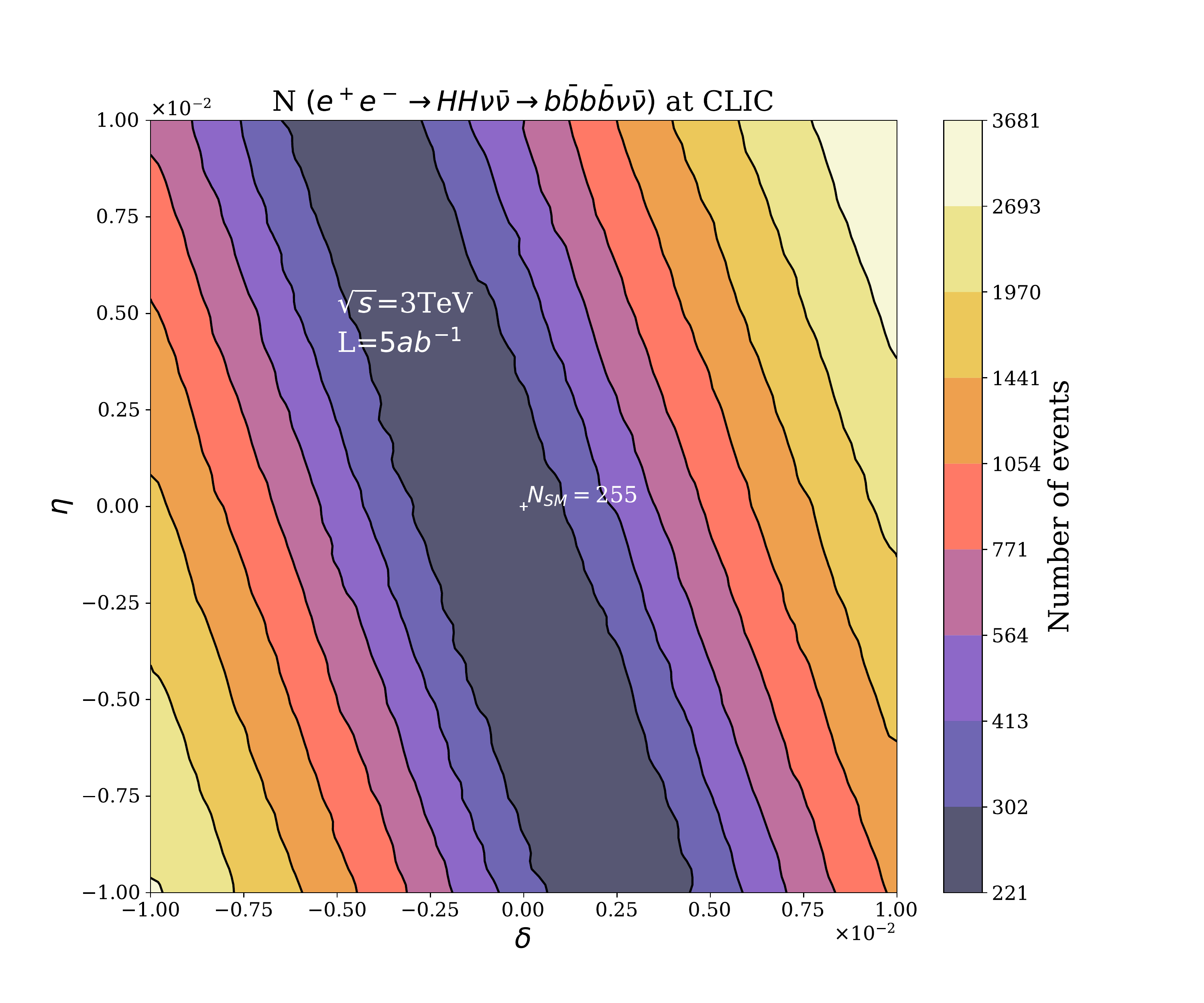}  \\
		\end{tabular}
\caption{EFT predictions for the number of events $N(e^+e^- \to HH \nu \bar \nu \to b \bar b b \bar b  \nu \bar \nu)$   in the 
$(\delta,\eta)$ plane for ILC (left panel) and CLIC (right panel).  Here the cuts and efficiencies described in the text are applied.  The SM predictions are also included}. 
\label{plot-NHEFT}
	\end{center}
\end{figure}
 
 Finally, to complete this study of the BSM Higgs physics with the use of the EWA, we have also computed the differential cross section for $e^+e^-\to HH \nu \bar \nu$ with respect to the invariant mass $M_{HH}$ of the di-Higgs pair as a function of the two relevant EFT parameters,  $\eta$ and $\delta$. The interest of this distribution is clear, since that departures of the BSM with respect to the SM predictions in the large $M_{HH}$ region precisely reflect the departures of the cross section at the subprocess $WW \to HH$ level with respect to the SM ones in the large subprocess $s$ region.  In \figref{dif-xsection} the two predictions of $ \sigma(WW \to HH)$ as a function of the subprocess energy $\sqrt{ s}$ (left) and $\frac{d\sigma(e^+e^- \to HH \nu \bar \nu)}{dM_{HH}}$ as a function of $M_{HH}$ (right) are displayed together, for various values of the relevant EFT parameters,  to show this correlation.  We also learn from this figure that a more detailed study of this enhancement in the tails of the distribution rates for $M_{HH}$ values at the TeV region could be used to improve the experimental sensitivity to these $\delta$ and $\eta$ parameters at the future $e^+e^-$ colliders. 
 
 In the last part of this section, we present a devoted study of the sensitivity to these most relevant parameters,  $\eta$ and 
 $\delta$,  based on the analysis of the event rates for the production of 4 $b$-jets and missing energy,  via the dominant Higgs decay into $b \bar b$ pairs.  The full process considered now is $e^+e^- \to HH \nu \bar \nu \to b \bar b b \bar b \nu \bar \nu$.  We study  the two colliders cases,  CLIC with $\sqrt{s}=3\, {\rm{TeV}}$  and $L=5\,  {\rm ab}^{-1}$,  and ILC  with $\sqrt{s}=1 \,{\rm {TeV}}$ and  $L=8\,  {\rm ab}^{-1}$.  We (naively) define the $b$-jets at the parton level and the missing energy  as that coming from the $\nu \bar \nu$ pairs.  We do not introduce detector simulation,  showering effects,  nor compute real backgrounds, thus this is rather a rough estimate of the sensitivity.  The following minimal detection cuts are applied to the final $b$-jets and missing energy variables:
 \begin{equation}
\label{cuts}
p_T^b>20\enspace \mathrm{GeV} \,\,\,\,;\,\,
\vert\eta^b\vert<2 \,\,\,\,;\,\,
\Delta R_{bb}>0.4 \,\,\,\,;\,\,E\!\!\!\!/_T>20\enspace \mathrm{GeV}
\end{equation}
 where,  $\Delta R_{bb}\equiv \sqrt{(\Delta \eta_{bb})^2+(\Delta \phi_{bb})^2}$. $\Delta\eta_{bb}$ and $\Delta\phi_{bb}$ are the separations in pseudorapidity and azimuthal angle of the two $b$-jets,   $p_T^b$ and $\eta^b$ are the transverse momentum and pseudorapidity of the $b$-jet and $E\!\!\!\!/_T$ is the transverse missing energy. These cuts are similar to those in Refs. \cite{Abramowicz:2016zbo, Gonzalez-Lopez:2020lpd}.  Obviously,  additional cuts more refined than these ones above could improve the acceptance in the ratio of the signal to background rates.   Particularly efficient could be requiring cuts on the invariant mass of the two $b$-jets pairs $M_{bb}$ to be close to the Higgs mass.   But,  for simplicity,  we keep our study just based on the above simplest/minimal  cuts. 
 
 For the generation of these events, and the computation of the cross section  
 with the above cuts implemented we employ MG5.  In addition to the reduction factors due to the Higgs decays,  with $BR^2\sim 0.58^2$, we have also applied the reduction factor$(\epsilon_b)^4$ due to the $b$-tagging efficiency, which we assume here to be $\epsilon_b \sim 0.8$.  The final predicted  cross sections with MG5 including all those cuts and reduction factors, 
 $\sigma(4b-{\rm jets} +E\!\!\!\!/_T)$,   are shown in \figref{plot-xsHEFT}.  The corresponding event rates are shown in \figref{plot-NHEFT}. The rates for ILC are displayed to the left,  and  for CLIC to the right,  in these figures.  The corresponding SM predictions are also included for comparison.  As we can see in both figures the rates from the BSM Higgs couplings are sufficiently large, compared to the SM ones,  to be detected,  both at ILC and CLIC,  if these effective couplings $\delta$ and $\eta$  are not too small,  i.e. if they are at the upper right and lower left regions of these plots.  For instance,  at ILC for  $(\delta, \eta)= (0.1, 0.1)$ we find 1087 events to be compared with 95 events in the SM; and at CLIC for  $(\delta, \eta)= (0.01, 0.01)$ we find 3681 events to be compared with 255 events in the SM.  These BSM rates are well separated from the SM rates and could be presumably tested at these colliders. 
 
\begin{figure}[t!]
	\begin{center}
		\begin{tabular}{cc}
			\centering
			\hspace*{-3mm}
			\includegraphics[scale=0.3]{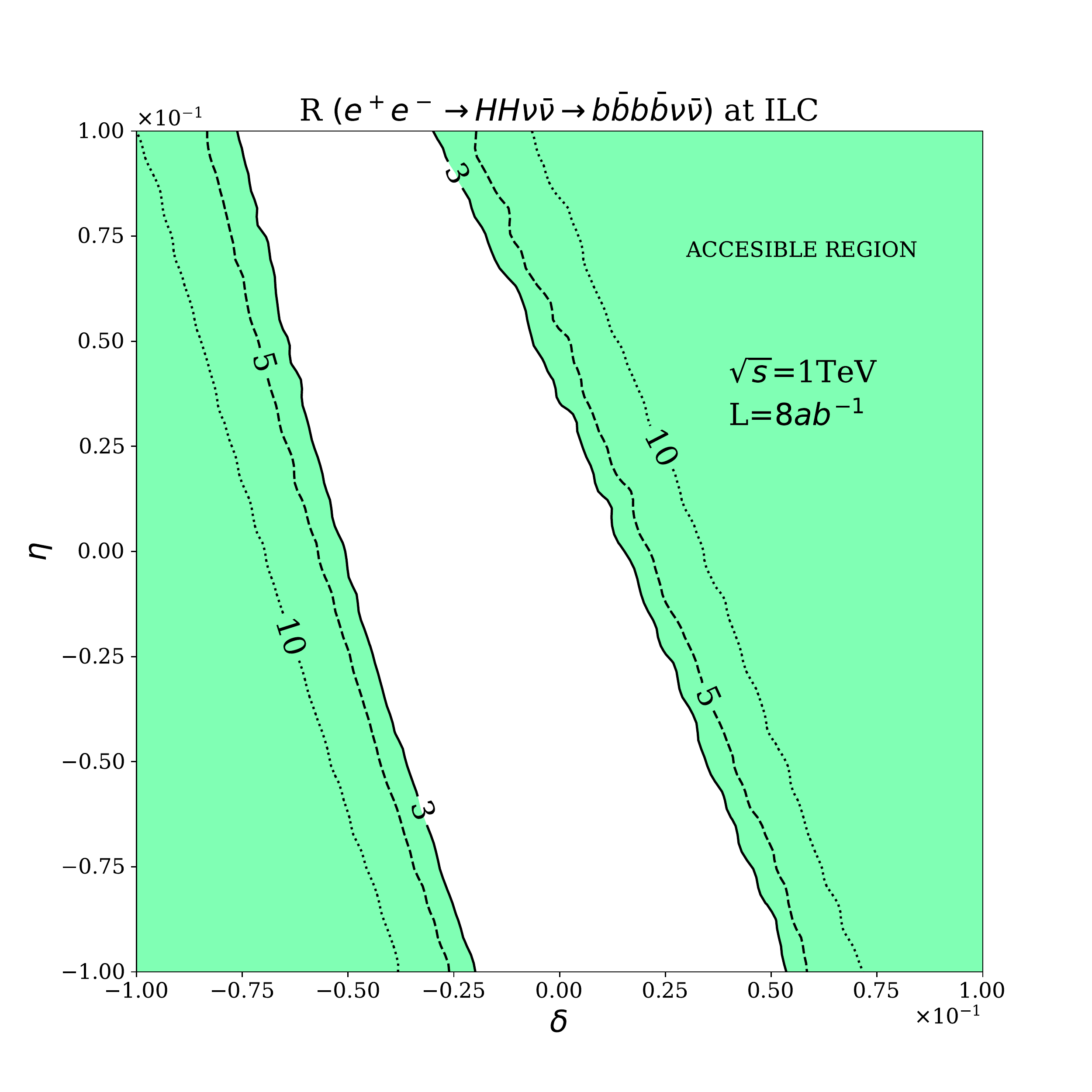} &	\includegraphics[scale=0.3]{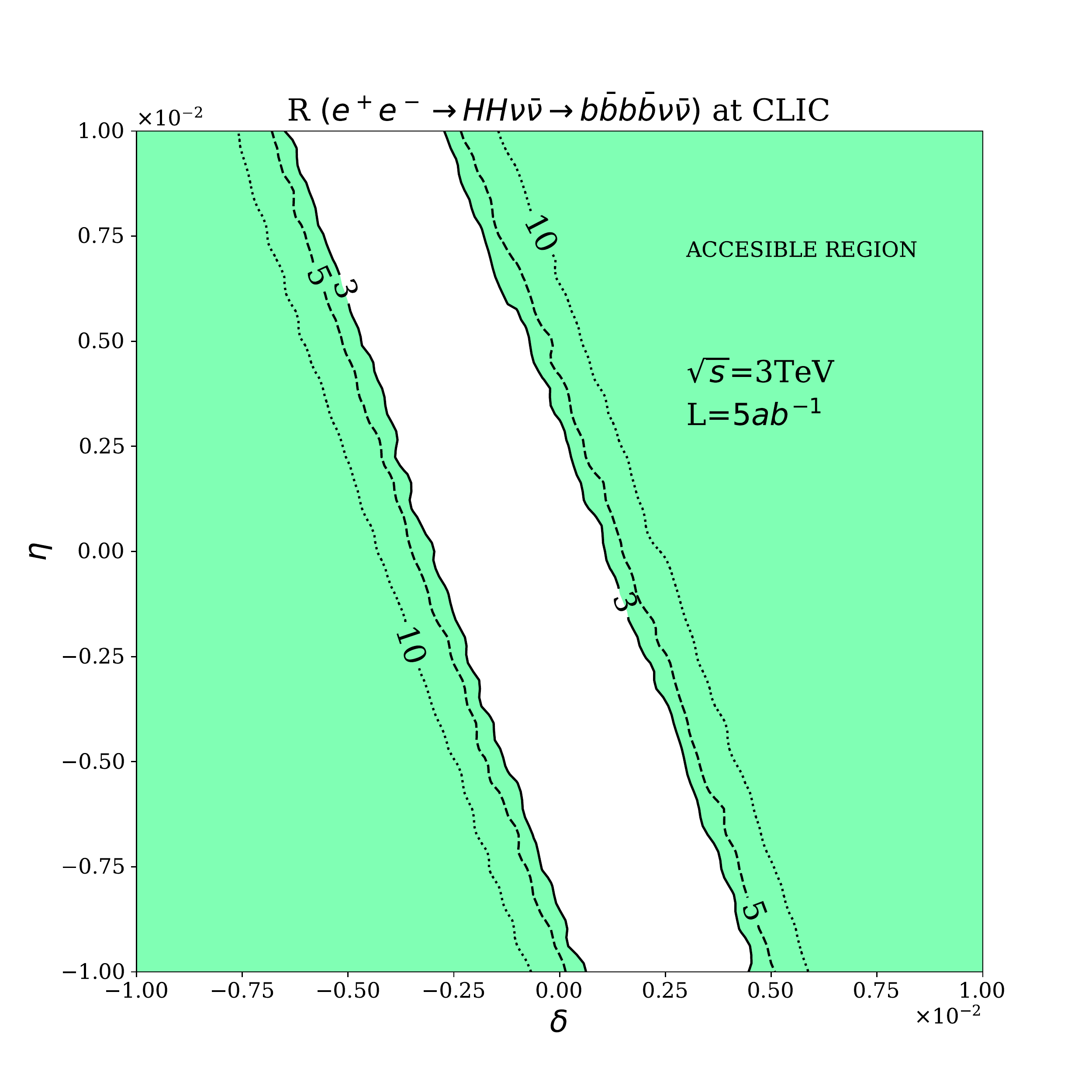}  \\
		\end{tabular}
\caption{Estimate of the potentially accessible regions in the $(\delta,\eta)$ plane based on the EFT prediction 
of the ratio $R$  defined in \eqref{Rdef},  for ILC (left panel) and CLIC (right panel).  Solid, dashed and dotted  contour-lines are for $R=3, 5, 10$ respectively.  Bright green areas correspond to $R>3$. }
\label{plot-RHEFT}
	\end{center}
\end{figure}
 
Finally, in order to quantify a bit better the sensitivity to these two EFT parameters, $\eta$ and $\delta$,   we have computed the following (theoretical) ratio $R$,  that is defined in terms of the previous mentioned event numbers for the BSM and SM cases,  $N_\mathrm{BSM}$ and $N_\mathrm{SM}$,  respectively,  by:
 \begin{equation}
 R=\frac{N_\mathrm{BSM}-N_\mathrm{SM}}
 {\sqrt{N_\mathrm{SM}}}
 \label{Rdef}
 \end{equation}
Our naive criterion of accessibility to a given parameter is set here in terms of the size of this ratio $R$.  We show in \figref{plot-RHEFT} the predictions (ILC at the left and CLIC at the right)  for the contour lines in the $(\delta, \eta)$ plane corresponding to $R=3$ (solid lines), $R=5$ (dashed lines) and $R=10$ (dotted lines).  Thus,  our conclusions on the accessible regions to these two parameters, $\delta$ and $\eta$ can be immediately extracted from this plot, depending on the required minimum $R$ value.  For instance, by requiring $R>3$ the areas in bright green are our estimates of the accessible regions to these $\delta$ and $\eta$ parameters.  It is also clear from this figure that the accessible regions at CLIC will be broader that at ILC,  as expected due to the higher energy. 
As we already stated, a detailed analysis taking into account all the backgrounds and the characteristics of the particular detectors at ILC and CLIC will be needed for a more precise conclusion,  but it is beyond the scope of this work and we  leave it for another future research.

\section{Conclusions}
\label{sec:conclus}

In this work we have studied in detail the scattering process $WW \to HH$ within the context of two EFTs: the HEFT and the SMEFT.  Both approaches parametrize in a very different way the possible departures from BSM Higgs physics with respect to the SM.  Within the HEFT the Higgs is a singlet field under the relevant EW and chiral symmetries, whereas in the SMEFT it is a component of a doublet together with the GBs of the EW symmetry breaking,  $SU(2)_L \times U(1)_Y \to U(1)_{\rm em}$.  The use of a linear (as in SMEFT) or a non-linear (as in HEFT)  representation for the GBs  may be more appropriate (or not),  for the study of BSM Higgs physics,  depending on the kind of dynamics underlying the UV fundamental theory that provides such EFT at lower energies.  If the underlying dynamics is strongly interacting the HEFT seems to be more appropriate, and the usual ordering of operators in the EChL by the increasing chiral dimension is the proper one that provides the hierarchy of the relevance of the EChL coefficients involved.  In contrast, the ordering in the relevance of the operators in the SMEFT is done in terms of the canonical dimension and therefore in terms of the inverse powers of the cut-off.

Through this work,   we have first presented,  in full detail,  the computation within the HEFT of the amplitude for this $WW \to HH$ scattering and evaluated the departures in the cross section with respect to the SM prediction as a function of the process energy $\sqrt{s}$, taking into account all the coefficients from the NLO EChL.   We have explored all the polarization channels   $W_XW_Y \to HH$ with $XY=LL, TT, LT, TL$, and also the total (unpolarized) cross section.  We have concluded that the $LL$ channel fully dominates the total cross section at the TeV energy domain,  and we have extracted the most relevant coefficients from the chiral dimension four HEFT Lagrangian.  These two coefficients,  $a_{dd\mV\mV1}$ and $a_{dd\mV\mV2}$,  have been identified with the usually called in the related literature,  $\eta$ and $\delta$ parameters, and correspond to the effective operators with four derivatives acting on the scalar fields.  The HEFT departures with respect to the SM  from these two coefficients can be large as summarized  in \figref{Total-Unpolarized}.  

Then we have also studied the case of SMEFT and we have identified which are the most relevant operators for this $WW \to HH$ scattering.  The corresponding predictions for the cross section of the various polarization channels and for various values of the relevant SMEFT Wilson coefficients were also provided.  The numerical results show that there are particular scenarios where the dim 8 operators with four derivatives acting on the scalar fields play an important role in those predictions.  We have identified these important dim 8 operators,  $\mathcal{O}_{\phi^4}^{(1,2,3)}$, and  have shown that for sizeable Wilson coefficients
 ${a_{\phi^4}^{(1,2,3)}} (1/{\Lambda^4}) \sim O(1) {\rm TeV}^{-4}$ they can provide relevant departures in the cross section with respect to the SM at the TeV energy domain.  These sizeable coefficient values may indicate the proximity to a strongly coupled theory.
 
We have also explored  in this work the consequences of doing the matching among the two EFTs,  HEFT and SMEFT,  at the level of the scattering amplitude, which is different than other approaches doing the matching at the Lagrangian level.  Proceeding with this matching of the two analytical predictions for the amplitudes 
$\mathcal{A}(WW\to HH)_{\rm HEFT}$ and $\mathcal{A}(WW\to HH)_{\rm SMEFT}$ and solving this matching equation in terms of the EFTs coefficients, we have arrived at the interesting relations among the coefficients of the two theories that are summarized in \eqref{eq:matching} and in \tabref{table:matching}. 

In the final part of this work,  we have explored the most relevant consequences of those departures found in $WW \to HH$, within  the EFT approach  to the BSM Higgs physics,  for the phenomenology of the planned $e^+e^-$ colliders at the TeV energy domain.  
Concretely, we have considered the two most energetic $e^+e^-$ colliders,  CLIC (3 TeV,  5 ab$^{-1}$) and ILC (1 TeV, 8 ab$^{-1}$).  In particular,  we have explored in detail the di-Higgs production  process $e^+e^- \to HH \nu \bar \nu$ which is shown here to proceed in the BSM case mainly via the $WW \to HH$ sub-process,  as it also occurs in the SM case.  In particular, we have shown numerically the dominance of the $W_LW_L \to HH$ subprocess in the total $e^+e^-$ cross section and the relevance of the two mentioned parameters $\delta$ and $\eta$ that provide the largest departures in  $\sigma(e^+e^- \to HH \nu \bar \nu)$ with respect to the SM one. 
In order, to conclude on the sensitivity to those two parameters at ILC and CLIC, we have studied the BSM rates for the case where the two final Higgs bosons decay to the most probable channels, i.e.  to $b \bar b$ pairs, leading to enhancements in the number of events with 4 $b$ jets plus missing energy with respect to the SM expected rates.  Studying the ratio of the BSM versus SM predictions by means of the variable $R$ defined in \eqref{Rdef} we have finally provided in \figref{plot-RHEFT} the potentially accessible regions  in the $(\delta, \eta)$ plane for both ILC and CLIC colliders.  These studies could improve significantly the sensitivity to these parameters and therefore also the knowledge about the underlying fundamental theory.

\section*{Acknowledgments}

The present work has received financial support from the ``Spanish Agencia 
Estatal de Investigaci\'on'' (AEI) and the EU
``Fondo Europeo de Desarrollo Regional'' (FEDER) through the project
   PID2019-108892RB-I00/AEI/10.13039/501100011033 and from the grant IFT
Centro de Excelencia Severo Ochoa SEV-2016-0597.  We also  acknowledge finantial support  from the European Union's Horizon 2020 research and innovation
programme under the Marie Sklodowska-Curie grant agreement No 674896 and
No 860881-HIDDeN,  he ITN ELUSIVES H2020-MSCA-ITN-2015//674896 and the RISE INVISIBLESPLUS H2020-MSCA-RISE-2015//690575. 
The work of RM is also supported by CONICET and ANPCyT under projects PICT 2016-0164, PICT 2017-2751 and PICT 2018-03682..
MR has received support from the European Union’s Horizon 2020 research and
innovation programme under the Marie Skłodowska-Curie grant agreement No 860881-
HIDDeN.



\bibliography{DHMR-arXiv-v1}

\end{document}